\documentclass[12pt,preprint]{aastex}
\usepackage{amsmath}
\usepackage{bm}
\usepackage{braket}% braket
\newcommand{\myemailTM}{tmiyatsu@ssu.ac.kr}
\newcommand{\myemailMKC}{cheoun@ssu.ac.kr}
\newcommand{\myemailKS}{koichi.saito@rs.tus.ac.jp}
\slugcomment{\today}
\shorttitle{EoS for neutron stars with hyperons and quarks}
\shortauthors{Miyatsu, Cheoun, \& Saito}
\begin{document}
\title{Equation of state for neutron stars with hyperons and quarks \\ in relativistic Hartree-Fock approximation}
\author{Tsuyoshi Miyatsu and Myung-Ki Cheoun}
\affil{Department of Physics, Soongsil University, Seoul 156-743, Korea}
\email{\myemailTM,\myemailMKC}
\and
\author{Koichi Saito}
\affil{Department of Physics, Faculty of Science and Technology, \\ Tokyo University of Science, Noda 278-8510, Japan}
\email{\myemailKS}
%%
% \author{Tsuyoshi Miyatsu\altaffilmark{1}, Myung-Ki Cheoun\altaffilmark{1}, and Koichi Saito\altaffilmark{2}}
% \email{\myemailTM}
% \altaffiltext{1}{Department of Physics, Soongsil University, Seoul 156-743, Korea}
% \altaffiltext{2}{Department of Physics, Faculty of Science and Technology, Tokyo University of Science, Noda 278-8510, Japan}
%%
\begin{abstract}
We construct the equation of state (EoS) for neutron stars explicitly including hyperons and quarks.
Using the quark-meson coupling model with relativistic Hartree-Fock approximation, the EoS for hadronic matter is derived by taking into account the strange ($\sigma^{\ast}$ and $\phi$) mesons as well as the light non-strange ($\sigma$, $\omega$, $\bm{\pi}$ and $\bm{\rho}$) mesons.
Relevant coupling constants are determined to reproduce the experimental data of nuclear matter and hypernuclei in SU(3) flavor symmetry.
For quark matter, we employ the MIT bag model with one-gluon-exchange interaction, and Gibbs criteria for chemical equilibrium in the phase transition from hadrons to quarks.
We find that the strange vector ($\phi$) meson and the Fock contribution make the hadronic EoS stiff, and that the maximum mass of a neutron star can be consistent with the observed mass of heavy neutron stars even if the coexistence of hadrons and quarks takes place in the core.
However, in the present calculation the transition to pure quark matter does not occur in stable neutron stars.
Furthermore, the lower bound of the critical chemical potential of the quark-hadron transition at zero temperature turns out to be around 1.5 GeV in order to be consistent with the recent observed neutron star data.
\end{abstract}
%%
%%%%%%%%%%%%%%%%%%%%%%%%%%%%%%%%%%%%%%%%%%%%%%%%%%%%%%%%%%%%%%%%%%%%%%%%%%%%%%%%%%%%%%%%%%%%%%%%%%%
\section{Introduction}
\label{sec:Introduction}
%%%%%%%%%%%%%%%%%%%%%%%%%%%%%%%%%%%%%%%%%%%%%%%%%%%%%%%%%%%%%%%%%%%%%%%%%%%%%%%%%%%%%%%%%%%%%%%%%%%

White dwarfs, neutron stars, and black holes, which are collectively referred to as compact stars, are consequent remnants of core-collapsing supernovae explosions.
Neutron stars may especially be believed to be cosmological laboratories for nuclear matter at extremely low temperature and high density, because the central density of neutron stars can reach several times higher than the normal nuclear density.
Thus, pulsar observations can provide some constraints on the equation of state (EoS) for dense nuclear matter \citep{Weber:1999qn,Glendenning:2000,Lattimer:2006xb}.
Of particular interest is the possibility of exotic degrees of freedom in the core of a neutron star, such as hyperons (Ys) \citep{Glendenning:1991es,Schaffner:1995th,Nishizaki:2002ih}, quark matter \citep{Itoh:1970uw,Witten:1984rs,Weber:2004kj}, some unusual condensations of boson-like matter \citep{Takatsuka:1978ku,Tatsumi:1988up,Glendenning:1998zx} and/or dark matter \citep{PerezGarcia:2010ap}.

Thanks to recent advances in astrophysical observations, we can obtain some precise information on the properties of neutron stars.
In particular, the discovery of massive neutron stars, PSR J1614-2230 with $1.97\pm0.04M_{\odot}$ \citep{Demorest:2010bx} and PSR J0348+0432 with $2.01\pm0.04M_{\odot}$ \citep{Antoniadis:2013pzd}, sets a strong constraint on the EoS for dense matter.
Meanwhile, at present, there have been many theoretical studies of the nuclear EoS based on many-body theories.
However, it is quite difficult to explain the heavy neutron stars by the EoS which have been calculated so far, if hyperons are supposed to exist in the core of neutron star, because the degrees of freedom of hyperons make the EoS very soft, and thus, the possible maximum mass of a neutron star is considerably reduced.

In order to solve this discrepancy between the observations and theories, which is the so-called hyperon puzzle, several useful approaches have been proposed in the last few years.
The relativistic Hartree-Fock (RHF) calculations with the quark-meson coupling (QMC) model \citep{Miyatsu:2011bc,Katayama:2012ge,Whittenbury:2013wma} and the Dirac-Brueckner-Hartree-Fock approach \citep{Katayama:2013zya,Katayama:2015dga} show remarkable results to settle the hyperon puzzle.
Their calculations indicate that the EoS with hyperons keeps stiffness even at high densities due to the suppression of hyperon production and a upward shift of the density at hyperon appearance.
In the relativistic mean-field (RMF) [or relativistic Hartree (RH)] calculations, the extension of SU(6) spin-flavor symmetry based on the quark model to SU(3) flavor symmetry with strange mesons is also helpful to clarify the problem in determining the couplings of the mesons to the octet baryons, because the strange vector ($\phi$) meson plays an important role in supporting massive neutron stars \citep{Weissenborn:2011ut,Miyatsu:2013yta,Lopes:2013cpa,Jiang:2012hy,Colucci:2013pya}.
It is also worth studying to understand the effect of hyperons in dense matter, introducing the interaction terms among  various mesons \citep{Bednarek:2011gd,Sulaksono:2012ny,Tsubakihara:2012ic} or using quantum Monte Carlo calculations \citep{Lonardoni:2014bwa}.
Furthermore, the modified $f(R)$ gravity may be another candidate for the solution of the hyperon puzzle \citep{Cheoun:2013tsa,Astashenok:2014pua}.

Although the phase transition from hadrons to quarks at high temperature and low baryon chemical potential is known to be crossover as explored by the first-principle lattice QCD simulation \citep{Aoki:2006we}, not only the order of the phase transition at zero temperature but also the existence of a critical end point in the QCD phase diagram are still unknown.
In dense matter such as a neutron star, the possibility of quark matter as well as hyperons may be expected to appear in the core and to influence a number of interesting astrophysical phenomena.
In general, the EoS for neutron stars with hadrons, leptons, and quarks, which is so-called hybrid stars, is considered by assuming a first-order phase transition \citep{Glendenning:1992vb,Glendenning:2001pe,Alford:2013aca}.
There have recently been many papers where the maximum mass of neutron stars with deconfined quarks can exceed $2M_{\odot}$ if the vector interaction between quarks is strong enough \citep{Bonanno:2011ch,Lenzi:2012xz,Logoteta:2013ipa,Logoteta:2013aca,Orsaria:2012je,Orsaria:2013hna,Klahn:2013kga,Yasutake:2014oxa}.
In addition, a smooth crossover based on the percolation picture to obtain the quark-hadron transition at zero temperature has been studied, and the EoS can also sustain the neutron-star mass of $2M_{\odot}$ using strongly interacting quark matter \citep{Masuda:2012kf,Masuda:2012ed}.
As explained in Refs. \citep{Lenzi:2012xz,Orsaria:2013hna}, without the strong vector interaction between quarks, it would be hard to explain the observation of compact stars with mass greater than around $2M_{\odot}$.

In this paper, we construct the EoS for neutron stars which can satisfy the $2M_{\odot}$ constraint from the recent astrophysical observations \citep{Demorest:2010bx,Antoniadis:2013pzd}, even when the exotic possibility such as hyperons and quarks are considered in the core.
Then, we study the properties of neutron stars, for instance, the mass, radius, and particle fractions, especially by focusing our mind on the effect of strangeness in neutron stars.

For uniform hadronic matter, we employ our previous EoS which have been calculated using the chiral quark-meson coupling (CQMC) model within RHF approximation \citep{Miyatsu:2013hea}.
The CQMC model is an extended version of the QMC model, in which the quark-quark hyperfine interactions caused by the one-gluon and pion exchanges are included.
Such hyperfine interactions play an impotent role in the baryon spectra in matter \citep{Nagai:2008ai,Miyatsu:2010zz,Saito:2010zw}.

In the QMC model, the quark mass in nuclear matter is reduced from the value in vacuum because of the condensed scalar ($\sigma$ and $\sigma^{\ast}$) fields depending on the nuclear density.
The decrease of the quark mass then leads to the variation of baryon internal structures at the quark level.
Such effects are considered self-consistently in the QMC model \citep{Guichon:1987jp,Saito:1994ki}.
In fact, the evidence for the medium modification of nucleon ($N$) structure in a nucleus has been observed in polarization transfer measurements in the quasi-elastic ($e,e^{\prime}p$) reaction at the Thomas Jefferson Laboratory, and the result supports the prediction of the QMC model \citep{Brooks:2011sa}.
This model has been successfully applied to the European Muon Collaboration effect \citep{Geesaman:1995yd,Saito:1993yw,Cloet:2006bq} and to explain various properties of finite nuclei as well as infinite nuclear matter \citep{Guichon:1995ue,Saito:1996sf,Saito:1996yb,Saito:2007rv}.
This approach can also be useful to evaluate the density dependence of various form factors \citep{Cheoun:2013exf,Cheoun:2013kla}.

In addition to the CQMC model, relevant coupling constants are determined so as to reproduce the experimental data of nuclear matter and hypernuclei in SU(3) flavor symmetry, including the hidden strange ($\sigma^{\ast}$ and $\phi$) mesons \citep{Weissenborn:2011ut,Miyatsu:2013yta,Lopes:2013cpa}.

In order to study how quark matter affects the neutron-star properties, we adopt the MIT bag model {\it without} the strong vector interaction between quarks, that is contrary to the recent calculations {\it with} it \citep{Bonanno:2011ch,Lenzi:2012xz,Logoteta:2013aca,Logoteta:2013ipa,Orsaria:2012je,Orsaria:2013hna,Klahn:2013kga,Yasutake:2014oxa}.
Furthermore, we assume the first-order phase transition from hadrons to quarks under Gibbs criteria \citep{Glendenning:1992vb,Glendenning:2001pe}.

This paper is organized as follows.
In Section \ref{sec:hadronic-matter}, a brief review of the formalism for Hartree-Fock calculation based on quantum hadrodynamics (QHD) \citep{Serot:1984ey} is presented.
The description of quark matter with one-gluon exchange effect using the MIT bag model and the phase transition from hadronic matter to quark matter are explained in Section \ref{sec:quark-matter} and \ref{sec:NS-matter}, respectively.
Numerical results and discussions are addressed in Section \ref{sec:result}.
Finally, we present a summary in Section \ref{sec:Summary}.

%%%%%%%%%%%%%%%%%%%%%%%%%%%%%%%%%%%%%%%%%%%%%%%%%%%%%%%%%%%%%%%%%%%%%%%%%%%%%%%%%%%%%%%%%%%%%%%%%%%
\section{Description of hadronic matter }
\label{sec:hadronic-matter}
%%%%%%%%%%%%%%%%%%%%%%%%%%%%%%%%%%%%%%%%%%%%%%%%%%%%%%%%%%%%%%%%%%%%%%%%%%%%%%%%%%%%%%%%%%%%%%%%%%%

We present the formulations for describing uniform hadronic matter.
In Quantum Hadrodynamics (QHD) \citep{Serot:1984ey}, the baryons are treated as point-like objects, and interact via the exchanges of mesons. On the other hand, we want to include the effect of baryon-structure variation in matter using the chiral quark-meson coupling (CQMC) model \citep{Nagai:2008ai,Miyatsu:2010zz,Saito:2010zw}.
In this paper, the Lagrangian density for hadronic matter is thus chosen to be
\begin{equation}
  \mathcal{L}_{H}=\mathcal{L}_{B}+\mathcal{L}_{M}+\mathcal{L}_{\rm int},
\end{equation}
where
\begin{equation}
  \mathcal{L}_{B}
  = \sum_{B}\bar{\psi}_{B}\left(i\gamma_{\mu}\partial^{\mu}-M_{B}\right)\psi_{B},
  \label{eq:Lagrangian-baryon}
\end{equation}
with $\psi_{B}$ being the baryon field and $M_{B}$ being the baryon mass in a vacuum.
The sum $B$ runs over the octet baryons: proton ($p$), neutron ($n$), $\Lambda$, $\Sigma^{+0-}$, and $\Xi^{0-}$.
For the free baryon masses, we take $M_{N}=939$ MeV,  $M_{\Lambda}=1116$ MeV,  $M_{\Sigma}=1193$ MeV,  and $M_{\Xi}=1318$ MeV, respectively.
Lepton Lagrangian is introduced in Section \ref{sec:NS-matter}.

In the present calculation, we study the effects of direct and exchange contributions on hadronic matter through not only the exchanges of non-strange mesons ($\sigma$, $\omega$, $\bm{\pi}$, and $\bm{\rho}$) but also those of strange mesons ($\sigma^{\ast}$ and $\phi$).
Thus, the meson term reads
\begin{align}
  \mathcal{L}_{M}
  &= \frac{1}{2}\left(\partial_{\mu}\sigma\partial^{\mu}\sigma-m_{\sigma}^{2}\sigma^{2}\right)
  +  \frac{1}{2}\left(\partial_{\mu}\sigma^{\ast}\partial^{\mu}\sigma^{\ast}-m_{\sigma^{\ast}}^{2}\sigma^{\ast2}\right)
  \nonumber \\
  &+ \frac{1}{2}m_{\omega}^{2}\omega_{\mu}\omega^{\mu} - \frac{1}{4}W_{\mu\nu}W^{\mu\nu}
  +  \frac{1}{2}m_{\phi}^{2}\phi_{\mu}\phi^{\mu} - \frac{1}{4}P_{\mu\nu}P^{\mu\nu}
  \nonumber \\
  &+ \frac{1}{2}m_{\rho}^{2}\bm{\rho}_{\mu}\cdot\bm{\rho}^{\mu} - \frac{1}{4}\bm{R}_{\mu\nu}\cdot\bm{R}^{\mu\nu}
  +  \frac{1}{2}\left(\partial_{\mu}\bm{\pi}\cdot\partial^{\mu}\bm{\pi}-m_{\pi}^{2}\bm{\pi}^{2}\right),
  \label{eq:Lagrangian-meson}
\end{align}
with
\begin{align}
  W_{\mu\nu} = \partial_{\mu}\omega_{\nu} - \partial_{\nu}\omega_{\mu},
  \label{eq:W-func} \\
  P_{\mu\nu} = \partial_{\mu}\phi_{\nu} - \partial_{\nu}\phi_{\mu},
  \label{eq:P-func} \\
  \bm{R}_{\mu\nu} = \partial_{\mu}\bm{\rho}_{\nu} - \partial_{\nu}\bm{\rho}_{\mu},
  \label{eq:R-func}
\end{align}
where the meson masses are respectively chosen as $m_{\sigma}=550$ MeV, $m_{\sigma^{\ast}}=975$ MeV, $m_{\omega}=783$ MeV, $m_{\phi}=1020$ MeV, $m_{\rho}=770$ MeV, and $m_{\pi}=138$ MeV.

The interaction Lagrangian is given by
\begin{align}
  \mathcal{L}_{\rm int}
  &= \sum_{B}\bar{\psi}_{B} \biggl[ g_{\sigma B}\left(\sigma\right)\sigma
  +  g_{\sigma^{\ast}B}\left(\sigma^{\ast}\right)\sigma^{\ast}
  -  g_{\omega B}\gamma_{\mu}\omega^{\mu}
  +  \frac{f_{\omega B}}{2\mathcal{M}}\sigma_{\mu\nu}\partial^{\nu}\omega^{\mu}
  \nonumber \\
  &- g_{\phi B}\gamma_{\mu}\phi^{\mu}
  +  \frac{f_{\phi B}}{2\mathcal{M}}\sigma_{\mu\nu}\partial^{\nu}\phi^{\mu}
  -  g_{\rho B}\gamma_{\mu}\bm{\rho}^{\mu}\cdot\bm{I}_{B}
  +  \frac{f_{\rho B}}{2\mathcal{M}}\sigma_{\mu\nu}\partial^{\nu}\bm{\rho}^{\mu}\cdot\bm{I}_{B}
  -  \frac{f_{\pi B}}{m_{\pi}}\gamma_{5}\gamma_{\mu}\partial^{\mu}\bm{\pi}\cdot\bm{I}_B \biggr] \psi_{B},
  \label{eq:Lagrangian-interaction}
\end{align}
where the common mass scale, $\mathcal{M}$, is taken to be the free nucleon mass, and $\bm{I}_B$ is the isospin matrix for baryon $B$.
The $\sigma$-, $\sigma^{\ast}$-, $\omega$-, $\phi$-, $\rho$-, $\pi$-$B$ coupling constants are respectively denoted by $g_{\sigma B}(\sigma)$, $g_{\sigma^{\ast}B}(\sigma^{\ast})$, $g_{\omega B}$, $g_{\phi B}$, $g_{\rho B}$ and $f_{\pi B}$, while $f_{\omega B}$, $f_{\phi B}$ and $f_{\rho B}$ are the tensor coupling constants for the vector mesons.
In the CQMC model, the coupling constants, $g_{\sigma B}(\sigma)$ and $g_{\sigma^{\ast}B}(\sigma^{\ast})$, have the scalar-field dependence which reflects the variation in the internal (quark) structure of baryons in matter, and they are caused by the attractive interactions due to the $\sigma$ and $\sigma^{\ast}$ exchanges.
For simplicity,  we adopt the following simple parametrizations for those coupling constants \citep{Miyatsu:2013yta,Miyatsu:2014wca,Tsushima:1997cu}:
\begin{align}
  g_{\sigma B}(\sigma)
	&= g_{\sigma B}b_{B}\left[1-\frac{a_{B}}{2}\left(g_{\sigma N}\sigma\right)\right],
	\label{eq:cc-sigma} \\
	g_{\sigma^{\ast}B}(\sigma^{\ast})
	&= g_{\sigma^{\ast}B}b_{B}^{\prime}
	\left[1-\frac{a_{B}^{\prime}}{2}\left(g_{\sigma^{\ast}\Lambda}\sigma^{\ast}\right)\right],
	\label{eq:cc-sigma-star}
\end{align}
where $g_{\sigma N}$ and $g_{\sigma^{\ast}\Lambda}$ are respectively the $\sigma$-$N$ and $\sigma^{\ast}$-$\Lambda$ coupling constants at zero density.
The effect of the variation of baryon structure at the quark level can be described with the parameters $a_{B}$ and $a_{B}^{\prime}$.
In addition, the extra parameters, $b_{B}$ and $b_{B}^{\prime}$, are necessary to express the effect of hyperfine interaction between two quarks \citep{Nagai:2008ai,Miyatsu:2010zz,Saito:2010zw}.
The couplings in the CQMC model are invariant under Lorentz transformation because they are functions of the scalar fields, and the values of the four parameters given in Equations \eqref{eq:cc-sigma} and \eqref{eq:cc-sigma-star} are tabulated in Table \ref{tab:CQMC-parameter}.
If we set $a_{B}=0$ and $b_{B}=1$, $g_{\sigma B}(\sigma)$ becomes identical to the $\sigma$-$B$ coupling constant in QHD. This is also true of the coupling $g_{\sigma^{\ast}B}(\sigma^{\ast})$.

In mean-field approximation, the meson fields are replaced by the constant mean-field values: $\bar{\sigma}$, $\bar{\sigma}^{\ast}$, $\bar{\omega}$, $\bar{\phi}$, and $\bar{\rho}$ (the $\rho^{0}$ field).
The mean-field value of the pion vanishes in the relativistic Hartree (RH) calculation, while the pion effect should be included in the relativistic Hartree-Fock (RHF) calculation, where the exchange contribution as well as the direct one are taken into account.
To sum up all orders of the tadpole (Hartree) and exchange (Fock) diagrams in the baryon Green's function, $G_{B}$, we use the Dyson's equation
\begin{equation}
  G_{B}(k) = G_{B}^{0}(k) + G_{B}^{0}(k)\Sigma_{B}(k)G_{B}(k),
	\label{eq:Dyson-equation}
\end{equation}
where $k^\mu$ is the four momentum of baryon, $\Sigma_{B}$ is the baryon self-energy, and $G_{B}^{0}$ is the Green's function in free space.
The baryon self-energy in matter is generally written as \citep{Serot:1984ey}
\begin{equation}
  \Sigma_{B}(k)
  = \Sigma_{B}^{s}(k)
  - \gamma_{0}\Sigma_{B}^{0}(k)
	+ (\bm{\gamma}\cdot\hat{k})\Sigma_{B}^{v}(k),
	\label{eq:baryon-self-engy}
\end{equation}
with $\hat{k}$ being the unit vector along the (three) momentum $\bm{k}$ and $\Sigma_{B}^{s(0)[v]}$ being the scalar part (the time component of the vector part) [the space component of the vector part] of the self-energy.
Therefore, the effective baryon mass, momentum, and energy in matter are respectively defined by including the self-energy in matter as follows \citep{Bouyssy:1987sh,Miyatsu:2011bc,Katayama:2012ge}
\begin{align}
  &M_{B}^{\ast}(k)
  = M_{B} + \Sigma_{B}^{s}(k),
  \label{eq:auxiliary-quantity-mass} \\
	&k_{B}^{\ast\mu}
	= (k_{B}^{\ast0},\bm{k}_{B}^{\ast}) = (k^{0}+\Sigma_{B}^{0}(k),\bm{k}+\hat{k}\Sigma_{B}^{v}(k)),
	\label{eq:auxiliary-quantity-momentum} \\
	&E_{B}^{\ast}(k)
	= \left[\bm{k}_{B}^{\ast2}+M_{B}^{\ast2}(k)\right]^{1/2}.
	\label{eq:auxiliary-quantity-energy}
\end{align}

The baryon self-energies in Equation (\ref{eq:baryon-self-engy}) are then calculated by \citep{Bouyssy:1987sh,Miyatsu:2011bc,Katayama:2012ge}
\begin{align}
  \Sigma_{B}^{s}(k)
  &= -g_{\sigma B}(\bar{\sigma})\bar{\sigma} - g_{\sigma^{\ast}B}(\bar{\sigma}^{\ast})\bar{\sigma}^{\ast}
  \nonumber \\
	&+ \sum_{B^{\prime},i}\frac{(I_{BB^{\prime}}^{i})^{2}}{(4\pi)^{2}k}\int_{0}^{k_{F_{B^{\prime}}}} dq \, q
	   \left[\frac{M_{B^{\prime}}^{\ast}(q)}{E_{B^{\prime}}^{\ast}(q)}B_{i}(k,q)
	+  \frac{q_{B^{\prime}}^{\ast}}{2E_{B^{\prime}}^{\ast}(q)}D_{i}(q,k)\right],
	\label{eq:BSE-scalar} \\
	\Sigma_{B}^{0}(k)
	&= -g_{\omega B}\bar{\omega} -g_{\phi B}\bar{\phi} - g_{\rho B}(\bm{I}_{B})_{z}\bar{\rho}
	-  \sum_{B^{\prime},i}\frac{(I_{BB^{\prime}}^{i})^{2}}{(4\pi)^{2}k}\int_{0}^{k_{F_{B^{\prime}}}}dq \, q A_{i}(k,q),
	\label{eq:BSE-time} \\
	\Sigma_{B}^{v}(k)
	&= \sum_{B^{\prime},i}\frac{(I_{BB^{\prime}}^{i})^2}{(4\pi)^{2}k}\int_{0}^{k_{F_{B^{\prime}}}}dq \, q
	   \left[\frac{q_{B^{\prime}}^{\ast}}{E_{B^{\prime}}^{\ast}(q)}C_{i}(k,q)
	+  \frac{M_{B^{\prime}}^{\ast}(q)}{2E_{B^{\prime}}^{\ast}(q)}D_{i}(k,q)\right],
	\label{eq:BSE-vector}
\end{align}
where $k_{F_B}$ is the Fermi momentum for baryon $B$, and the factor, $I_{BB^{\prime}}^{i}$, is the isospin weight at the meson-$BB^\prime$ vertex in the Fock diagram.

In order to include the effect of the finite size of baryons, a form factor at each interaction vertex should be introduced, and, in the present calculation, we employ a dipole-type form factor \citep{Katayama:2012ge}:
\begin{equation}
  F_{i}({p}^{2})
  = \frac{1}{\left(1- p^{2}/\Lambda_{i}^{2}\right)^{2}},
  \label{eq:form-factor}
\end{equation}
where $p^\mu$ is the (four) momentum transfer, $\Lambda_{i}$ is a cutoff parameter, and $i$ specifies the interaction vertex (see the 1st column of Table \ref{tab:BSE}).
In the interaction Lagrangian density, we then replace all coupling constants with those multiplied by the form factor.
In fact, the effect of form factor can not be seen at the Hartree level, because the momentum transfer in the meson exchange between two baryons vanishes.
However, because the exchanged momentum can be finite in the Fock term, it may become significant as the density increases.
In addition, the retardation effect in the Fock term are ignored, since it gives at most a few percent contribution to the baryon self-energy \citep{Serot:1984ey,Katayama:2012ge,Whittenbury:2013wma}.
The functions $A_{i}$, $B_{i}$, $C_{i}$, and $D_{i}$ in Equations \eqref{eq:BSE-scalar}--\eqref{eq:BSE-vector} are explicitly given in Table \ref{tab:BSE}, in which the following functions are used \citep{Katayama:2012ge}:
\begin{align}
  \Theta_{i}(k,q)
  &= \frac{\Lambda_{i}^{8}}{(m_{i}^{2}-\Lambda_{i}^{2})^{4}}
     \left(\ln\left[\frac{M_i^{+}(k,q)}{M_i^{-}(k,q)}\frac{L_i^{-}(k,q)}{L_i^{+}(k,q)}\right]
	+  \sum_{n=1}^{3}\left(m_{i}^{2}-\Lambda_{i}^{2}\right)^{n}N_{i}^{n}(k,q)\right),
	\label{eq:Theta-func} \\
	\Phi_{i}(k,q)
	&= \frac{1}{4kq}\left[\left(k^{2}+q^{2}+m_{i}^{2}\right)\Theta_{i}(k,q) - \Lambda_{i}^{8}N_{i}^{3}(k,q)\right],
	\label{eq:Phi-func} \\
	\Psi_{i}(k,q)
	&= \left(k^{2}+q^{2}-m_{i}^{2}/2\right)\Phi_{i}(k,q) -kq\Theta_{i}(k,q) + \Omega_{i}(k,q),
	\label{eq:Psi-func} \\
	\Pi_{i}(k,q)
	&= \left(k^{2}+q^{2}\right)\Phi_{i}(k,q) - kq\Theta_{i}(k,q) + \Omega_{i}(k,q),
	\label{eq:Pi-func} \\
	\Gamma_{i}(k,q)
	&= \left[k\Theta_{i}(k,q)-2q\Phi_{i}(k,q)\right],
	\label{eq:Gamma-func}
\end{align}
where
\begin{align}
  \Omega_{i}(k,q)
  &= \frac{\Lambda_{i}^{8}}{4kq}\left[N_{i}^{2}(k,q)+\left(k^{2}+q^{2}+\Lambda_{i}^{2}\right)N_{i}^{3}(k,q)\right],
  \label{eq:Omega-func}  \\
	L_{i}^{\pm}(k,q)
	&= \Lambda_{i}^{2} + (k \pm q)^{2},
	\label{eq:L-func} \\
	M_{i}^{\pm}(k,q)
	&= m_{i}^{2} + (k \pm q)^{2},
	\label{eq:M-func} \\
	N_{i}^{n}(k,q)
	&= \frac{(-1)^{n}}{n}\left(\left[L_{i}^{+}(k,q)\right]^{-n}-\left[L_{i}^{-}(k,q)\right]^{-n}\right).
	\label{eq:N-func}
\end{align}
We note that the effect of form factor vanishes in the limit $\Lambda_{i} \to \infty$.

By solving the Euler-Lagrange equations for the meson fields, as usually adopted in the relativistic mean-field approach, the mean-field values of $\bar{\sigma}$, $\bar{\sigma}^{\ast}$, $\bar{\omega}$, $\bar{\phi}$, and $\bar{\rho}$ in Equations \eqref{eq:BSE-scalar} and \eqref{eq:BSE-time} are given by
\begin{align}
  \bar{\sigma}
  &= \sum_{B}\frac{g_{\sigma B}}{m_{\sigma}^{2}}C_{B}(\bar{\sigma})\rho_{B}^{s},
  \label{eq:EOM-sigma} \\
  \bar{\sigma}^{\ast}
  &= \sum_{B}\frac{g_{\sigma^{\ast}B}}{m_{\sigma^{\ast}}^{2}}C_{B}^{\prime}(\bar{\sigma}^{\ast})\rho_{B}^{s},
  \label{eq:EOM-sigma-star} \\
  \bar{\omega}
  &= \sum_{B}\frac{g_{\omega B}}{m_{\omega}^{2}}\rho_{B},
  \label{eq:EOM-omega} \\
  \bar{\phi}
  &= \sum_{B}\frac{g_{\phi B}}{m_{\phi}^{2}}\rho_{B},
  \label{eq:EOM-phi} \\
  \bar{\rho}
  &= \sum_{B}\frac{g_{\rho B}}{m_{\rho}^{2}}(\bm{I}_{B})_{3}\rho_{B},
  \label{eq:EOM-rho}
\end{align}
where the scalar density, $\rho_{B}^{s}$, and the baryon number density, $\rho_{B}$, read
\begin{align}
  \rho_{B}^{s}
  &= \frac{2J_{B}+1}{2\pi^{2}}\int_{0}^{k_{F_{B}}} dk \, k^{2} \frac{M_{B}^{\ast}(k)}{E_{B}^{\ast}(k)},
  \label{eq:baryon-scalar-density} \\
  \rho_{B}
  &= \frac{2J_{B}+1}{2\pi^{2}}\int_{0}^{k_{F_{B}}} dk \, k^{2}
  =  \frac{2J_{B}+1}{6\pi^{2}}{k^{3}_{F_{B}}},
  \label{eq:baryon-number-density}
\end{align}
with $J_{B}$ being the spin degeneracy factor of baryon $B$.
In Equations \eqref{eq:EOM-sigma} and \eqref{eq:EOM-sigma-star}, $C_{B}(\bar{\sigma})$ and $C_{B}^{\prime}(\bar{\sigma}^{\ast})$ are respectively  the scalar polarizabilities at the $\sigma$-$B$ and $\sigma^{\ast}$-$B$ interactions.
In the CQMC model, they can be expressed by the following parametrizations \citep{Miyatsu:2013yta,Miyatsu:2014wca,Tsushima:1997cu}:
\begin{align}
  C_{B}(\bar{\sigma})
	&= b_{B} \left[ 1 - a_{B}\left(g_{\sigma N} \bar{\sigma} \right) \right],
	\label{eq:scalar-density-ratio-sigma} \\
	C_{B}^{\prime}(\bar{\sigma}^{\ast})
	&= b_{B}^{\prime} \left[ 1 - a_{B}^{\prime}\left(g_{\sigma^{\ast}\Lambda} \bar{\sigma}^{\ast} \right) \right],
	\label{eq:scalar-density-ratio-sigma-star}
\end{align}
where the parameters, $a_{B}$, $b_{B}$, $a_{B}^{\prime}$, and $b_{B}^{\prime}$, take the same values as in Equations \eqref{eq:cc-sigma} and \eqref{eq:cc-sigma-star}.

With a self-consistent calculation of the baryon self-energies in Equations \eqref{eq:BSE-scalar}--\eqref{eq:BSE-vector},  the energy density for hadronic matter, which includes the baryon and meson contributions, can be expressed as
\begin{equation}
  \epsilon_{H}
  = \sum_{B}\frac{2J_{B}+1}{(2\pi)^{3}}\int_{0}^{k_{F_{B}}}d\bm{k}\left[T_{B}(k)+\frac{1}{2}V_{B}(k)\right],
	\label{eq:hadron-engy-density}
\end{equation}
with
\begin{align}
  T_{B}(k)
  &= \frac{M_{B}M_{B}^{\ast}(k)+kk_{B}^{\ast}}{E_{B}^{\ast}(k)},
	\label{eq:hadron-engy-density-kinetic} \\
	V_{B}(k)
	&= \frac{M_{B}^{\ast}(k)\Sigma_{B}^{s}(k)+k_{B}^{\ast}\Sigma_{B}^{v}(k)}{E_{B}^{\ast}(k)}-\Sigma_{B}^{0}(k).
	\label{eq:hadron-engy-density-potential}
\end{align}
Then, the pressure for hadronic matter is given by
\begin{equation}
  P_{H} = n_{B}^{2}\frac{\partial}{\partial n_{B}}\left(\frac{\epsilon_{H}}{n_{B}}\right),
	\label{eq:hadron-pressure}
\end{equation}
where the total baryon number density is defined by $n_{B}=\sum_{B}\rho_{B}$.

%%%%%%%%%%%%%%%%%%%%%%%%%%%%%%%%%%%%%%%%%%%%%%%%%%%%%%%%%%%%%%%%%%%%%%%%%%%%%%%%%%%%%%%%%%%%%%%%%%%
\section{Quark matter description}
\label{sec:quark-matter}
%%%%%%%%%%%%%%%%%%%%%%%%%%%%%%%%%%%%%%%%%%%%%%%%%%%%%%%%%%%%%%%%%%%%%%%%%%%%%%%%%%%%%%%%%%%%%%%%%%%

We briefly present the description of uniform quark matter.
The thermodynamic potential can be simply expressed by \citep{Freedman:1977gz,Farhi:1984qu,Alcock:1986hz,Haensel:1986qb}
\begin{equation}
  \Omega = \sum_{q}\Omega_{q} + B,
  \label{eq:total-thermodynamic-potential}
\end{equation}
with the quark term, $\Omega_{q}$, and the bag constant, $B$.
The sum $q$ runs over three-flavor quarks ($u$, $d$ and $s$), and the quark thermodynamic potential is given by a sum of the kinetic term and the interaction term due to the one-gluon exchange (OGE) \citep{Freedman:1977gz}:
\begin{equation}
  \Omega_{q}
  = -\frac{g_{q}\left(2J_{q}+1\right)}{24\pi^{2}} \left[ F\left(\mu_{q},m_{q}\right)
  - \frac{2\alpha_{c}}{\pi}G\left(\mu_{q},m_{q}\right) \right],
  \label{eq:quark-thermodynamic-potential}
\end{equation}
where the color weight for quark species, $g_{q}$, is equal to 3, $J_{q}$ is the the spin degeneracy factor, and $\alpha_{c}$ is the QCD fine structure constant.
In addition, the first-order correction due to the gluon interaction is included in the second term of Equation \eqref{eq:quark-thermodynamic-potential}.
The functions $F$ and $G$ are written as
\begin{align}
  F\left(\mu_{q},m_{q}\right)
  &= \mu_{q}\sqrt{\mu_{q}^{2}-m_{q}^{2}}\left(\mu_{q}^{2}-\frac{5}{2}m_{q}^{2}\right)
  +  \frac{3}{2}m_{q}^{4}\ln\left(\frac{\mu_{q}+\sqrt{\mu_{q}^{2}-m_{q}^{2}}}{m_{q}}\right),
  \label{eq:F-func} \\
  G\left(\mu_{q},m_{q}\right)
  &= 3\left[ \mu_{q}\sqrt{\mu_{q}^{2}-m_{q}^{2}}
  -  m_{q}^{2}\ln\left(\frac{\mu_{q}+\sqrt{\mu_{q}^{2}-m_{q}^{2}}}{m_{q}}\right) \right]^{2}
  -  2\left(\mu_{q}^{2}-m_{q}^{2}\right)^{2},
  \label{eq:G-func}
\end{align}
with $\mu_{q}$ being the quark chemical potential and $m_{q}$ being the current quark mass.
We here take $m_{u(d)}=5$ MeV and $m_{s}=150$ MeV.

The quark number density is related to $\Omega_{q}$ via $\rho_{q}=-\partial\Omega_{q}/\partial\mu_{q}$, and the baryon density and the charge density in quark matter are given by $n_{B}=\frac{1}{3}\sum_{q}\rho_{q}$ and $n_{C}=\frac{2}{3}\rho_{u}-\frac{1}{3}\left(\rho_{d}+\rho_{s}\right)$, respectively.
The energy density and pressure for quark matter are then written as
\begin{align}
  \varepsilon_{Q}
  &= \sum_{q}\left(\Omega_{q}+\mu_{q}\rho_{q}\right) + B,
  \label{eq:total-engy-density-quark} \\
  P_{Q}
  &= -\sum_{q}\Omega_{q} - B.
  \label{eq:total-pressure-quark}
\end{align}

Instead of hadronic matter, quark matter is expected to exist at very high densities such as the center of a neutron star.
However, the exact value of the transition density from hadron phase to quark one is still unknown.
In the present calculation, in order to study the effect of quark matter on neutron stars, we introduce a density-dependent bag constant, which is assumed to be given by a gaussian parametrization \citep{Burgio:2001mk,Burgio:2002sn}
\begin{equation}
  B\left(n_{B}\right)
  = B_{\infty} + \left(B_{0}-B_{\infty}\right)\exp\left[-\beta\left(\frac{n_{B}}{n_{0}}\right)^{2} \right],
  \label{eq:DDbag}
\end{equation}
with $n_{0}$ and $\beta$ being the normal nuclear density and a parameter, respectively.
Since quark matter may appear at very high densities and the final results do not sensitively depend on $B_{0}$ and $B_{\infty}$, we here fix $B_{0}=400$ MeV fm$^{-3}$ and $B_{\infty}=50$ MeV fm$^{-3}$ \citep{Maieron:2004af,Maruyama:2007ey}.
Due to the density dependence, the bag constant in Equation \eqref{eq:total-engy-density-quark} becomes $B\left(n_{B}\right)$,  and that in Equation \eqref{eq:total-pressure-quark} should be replaced as 
\begin{equation}
  - B \to n_{B}\frac{dB\left(n_{B}\right)}{dn_{B}} - B\left(n_{B}\right).
  \label{eq:pressure-quark-DDbag}
\end{equation}

The coupling between a quark and gluon is scale-dependent, and the lowest-order coupling at momentum transfer $Q^{2}$ is given by $\alpha_{c}\left(Q^{2}\right)=12\pi/\left[\left(33-2N_{f}\right)\ln\left(Q^{2}/\Lambda_{\rm QCD}^{2}\right)\right]$ with $N_{f}$ being the number of quark flavors and $\Lambda_{\rm QCD}\simeq200$ MeV.
In practice, it can be parametrized in a convenient form \citep{Capstick:1986bm}
\begin{equation}
  \alpha_{c}\left(Q^{2}\right) = 0.25e^{-Q^{2}} + 0.15e^{-Q^{2}/10} + 0.20e^{-Q^{2}/1000},
  \label{eq:QCDCC}
\end{equation}
where $Q^{2}$ is in GeV$^{2}$.
The coupling in Equation \eqref{eq:QCDCC} shows $\alpha_{c}=0.6$ at zero momentum transfer, and the momentum transfer is replaced by the average of quark chemical potentials, $\overline{\mu^{2}}=\frac{1}{3}\left(\mu_{u}^{2}+\mu_{d}^{2}+\mu_{s}^{2}\right)$ \citep{Freedman:1977gz}.

%%%%%%%%%%%%%%%%%%%%%%%%%%%%%%%%%%%%%%%%%%%%%%%%%%%%%%%%%%%%%%%%%%%%%%%%%%%%%%%%%%%%%%%%%%%%%%%%%%%
\section{Neutron-star matter and phase transition}
\label{sec:NS-matter}
%%%%%%%%%%%%%%%%%%%%%%%%%%%%%%%%%%%%%%%%%%%%%%%%%%%%%%%%%%%%%%%%%%%%%%%%%%%%%%%%%%%%%%%%%%%%%%%%%%%

In order to calculate the properties of neutron-star matter, the charge neutrality and $\beta$ equilibrium under weak processes are imposed.
The leptons must be introduced to achieve these conditions, and the Lagrangian density can be written as
\begin{equation}
  \mathcal{L}_{\ell}
  = \sum_{\ell}\bar{\psi}_{\ell}\left(i\gamma_{\mu}\partial^{\mu}-m_{\ell}\right)\psi_{\ell},
  \label{eq:Lagrangian-lepton}
\end{equation}
with $\psi_{\ell}$ being the lepton field and $m_{\ell}$ being its mass.
The sum $\ell$ is for $e^{-}$ and $\mu^{-}$.
The lepton energy density, pressure, and number density are then given by
\begin{align}
  \epsilon_{\ell}
  &= \sum_{\ell}\frac{2J_{\ell}+1}{2\pi^{2}} \int_{0}^{k_{F_{\ell}}} dk \, k^{2} \sqrt{k^{2}+m_{\ell}^{2}},
  \label{eq:lepton-engy-density} \\
  P_{\ell}
  &= \frac{1}{3}\sum_{\ell}\frac{2J_{\ell}+1}{2\pi^{2}} \int_{0}^{k_{F_{\ell}}} dk \, \frac{k^{4}}{\sqrt{k^{2}+m_{\ell}^{2}}},
  \label{eq:lepton-pressure} \\
  \rho_{\ell}
  &= \frac{2J_{\ell}+1}{2\pi^{2}}\int_{0}^{k_{F_{\ell}}} dk \, k^{2}
  =  \frac{2J_{\ell}+1}{6\pi^{2}}{k^{3}_{F_{\ell}}},
  \label{eq:lepton-number-density}
\end{align}
where $J_{\ell}$ is the spin degeneracy factor of lepton $\ell$.
The total energy density and pressure for hadronic (quark) matter are given by the sum of the hadron (quark) and lepton parts, namely $\epsilon=\epsilon_{H(Q)}+\epsilon_{\ell}$ and $P=P_{H(Q)}+P_{\ell}$.
Furthermore, the condition of $\beta$ equilibrium is expressed as \citep{Glendenning:1992vb,Glendenning:2001pe,Maruyama:2007ey}
\begin{align}
  \mu_{n}
  &= \mu_{\Lambda} = \mu_{\Sigma^{0}} = \mu_{\Xi^{0}} = \mu_{u} + 2\mu_{d},
  \label{eq:beta-neutral} \\
  \mu_{n} + \mu_{e}
  &= \mu_{\Sigma^{-}} = \mu_{\Xi^{-}} = \mu_{d} = \mu_{s},
  \label{eq:beta-minus} \\
  \mu_{n} - \mu_{e}
  &= \mu_{p} = \mu_{\Sigma^{+}},
  \label{eq:beta-plus} \\
  \mu_{e}
  &= \mu_{\mu},
  \label{eq:beta-lepton}
\end{align}
with the chemical potentials for baryons, quarks, and leptons.
We note that the properties of neutron-star matter are generally characterized by two independent chemical potentials related to baryon number density and total charge density.

In order to describe the coexistence of hadrons and quarks, we impose Gibbs criterion for chemical equilibrium \citep{Glendenning:1992vb,Glendenning:2001pe}.
Under Gibbs criterion, in the mixed phase (MP), pressure in the hadron phase (HP) must balance with that in the quark phase (QP) to ensure mechanical stability as follows,
\begin{equation}
  P_{\rm HP}\left(\mu_{n},\mu_{e}\right) = P_{\rm QP}\left(\mu_{n},\mu_{e}\right).
  \label{eq:Gibbs-criterion-press}
\end{equation}
In the MP, where the condition, Equation \eqref{eq:Gibbs-criterion-press}, is satisfied, the charge neutrality can be expressed as
\begin{equation}
  (1-\chi)n_{C}^{\rm HP} + \chi n_{C}^{\rm QP} = 0,
  \label{eq:Gibbs-criterion-charge}
\end{equation}
with $n_{C}^{\rm HP}$ ($n_{C}^{\rm QP}$) beign the charge density in the HP (QP), and  $\chi$ being a volume fraction.
The total energy density and baryon number density in the MP are then given by
\begin{align}
  \epsilon_{\rm MP}
  &= (1-\chi)\epsilon_{\rm HP} + \chi\epsilon_{\rm QP},
  \label{eq:total-engy-density-mixed-phase} \\
  n_{B}^{\rm MP}
  &= (1-\chi)n_{B}^{\rm HP} + \chi n_{B}^{\rm QP}.
  \label{eq:baryon-number-density-mixed-phase}
\end{align}

%%%%%%%%%%%%%%%%%%%%%%%%%%%%%%%%%%%%%%%%%%%%%%%%%%%%%%%%%%%%%%%%%%%%%%%%%%%%%%%%%%%%%%%%%%%%%%%%%%%
\section{Numerical results}
\label{sec:result}
%%%%%%%%%%%%%%%%%%%%%%%%%%%%%%%%%%%%%%%%%%%%%%%%%%%%%%%%%%%%%%%%%%%%%%%%%%%%%%%%%%%%%%%%%%%%%%%%%%%

%%%%%%%%%%%%%%%%%%%%%%%%%%%%%%%%%%%%%%%%%%%%%%%%%%%%%%%%%%%%%%%%%%%%%
\subsection{Coupling constants and matter properties}
\label{subsec:result-matter}
%%%%%%%%%%%%%%%%%%%%%%%%%%%%%%%%%%%%%%%%%%%%%%%%%%%%%%%%%%%%%%%%%%%%%

Firstly, we consider the iso-symmetric nuclear matter around the normal nuclear density, $n_{0}=0.155$ fm$^{-3}$, in which region hyperons do not appear.
Then, the couplings for $g_{\sigma N}$, $g_{\omega N}$, and  $g_{\phi N}$ are determined so as to fit the saturation energy ($-16.1$ MeV) at $n_{0}$.
We note that, in SU(6) symmetry, the $\phi$ meson does not couple to the nucleon ($g_{\phi N}=0$), but that it does couple to the nucleon in SU(3) symmetry through the following relation:
\begin{equation}
  g_{\phi N}
	= \frac{\sqrt{3}z-\tan\theta_{v}}{1+\sqrt{3}z\tan\theta_{v}} g_{\omega N},
	\label{eq:SU3-mix}
\end{equation}
where the mixing angle and the ratio of a coupling for the octet state to one for the singlet state are respectively chosen to be $\theta_{v}=37.50^{\circ}$ and $z=0.1949$ \citep{Miyatsu:2013yta,Miyatsu:2014wca,Rijken:2010zzb}.
The coupling constants, $g_{\sigma N}$, $g_{\omega N}$, $g_{\phi N}$, and $g_{\rho N}$, are shown in Table \ref{tab:CCN}.
For comparison, we also show the results for the case in SU(6) spin-flavor symmetry.
The $\rho$-$N$ coupling constant, $g_{\rho N}$, is determined so as to fit the the symmetry energy, $E_{\rm sym}=32.5$ MeV \citep{Tsang:2012se,Lattimer:2012xj}.  
For the tensor coupling constants, we use the ratios, $f_{\omega N}/g_{\omega N}$, $f_{\phi N}/g_{\phi N}$, and $f_{\rho N}/g_{\rho N}$, which have been suggested by the ESC08 model \citep{Rijken:2010zzb}\footnote{The ESC08 model may at present be the most complete meson-exchange model based on SU(3) flavor symmetry, which can describe the $Y$-$N$ and $Y$-$Y$ as well as $N$-$N$ interactions.}.
Furthermore, the $\pi$-$N$ coupling constants, $f_{\pi N}$, are also chosen to be the value given in the same model.
In the present calculation, we assume that the $\sigma^{\ast}$ meson does not couple to the nucleon ($g_{\sigma^{\ast}N}=0$).

In Table \ref{tab:nuclear-properties}, we present the several properties of symmetric nuclear matter at $n_{0}$.
The symmetry energy, $E_{\rm sym}$, is defined in terms of the 2nd derivative of the total energy with respect to the difference between proton and neutron densities.
Then, the slope (curvature) parameter of the symmetry energy, $L_{0}$ ($K_{\rm sym}$), is evaluated by the 1st (2nd) derivative of $E_{\rm sym}$ with respect to the baryon density \citep{Chen:2009wv,Agrawal:2012rx,Sulaksono:2012ny}.
The incompressibility in the SU(3) case is slightly smaller than that in the SU(6) case, and it stands closer to the range of $K_{0}=240\pm20$ MeV, which is derived from the isoscalar giant monopole resonance \citep{Chen:2009wv}.
The slope parameters of the symmetry energy in both cases show the reasonable values predicted by theoretical calculations \citep{Danielewicz:2008jn,Tsang:2012se}.
Moreover, in both of the SU(6) and SU(3) symmetries, the 2nd derivative values of the isobaric incompressibility coefficient lie well within the theoretical analysis, which has recently been estimated to be $K_{{\rm sat},2}=-370\pm120$ MeV \citep{Chen:2009wv}.

Next, we study the coupling constants for hyperons.
The coupling constants for the $\sigma$ meson are chosen so as to give reasonable hyperon potentials \citep{Miyatsu:2013yta,Miyatsu:2014wca,Schaffner:1995th,Yang:2008am}.
Using the baryon self-energy given in Equations \eqref{eq:BSE-scalar} and \eqref{eq:BSE-time}, the potential for hyperon $Y$ embedded in the matter of baryon $B$ is expressed by the Schr\"{o}dinger-equivalent form \citep{Jaminon:1981xg}:
\begin{equation}
  U_{Y}^{(B)} = \Sigma_{Y}^{sH} - \Sigma_{Y}^{0H}
              + \frac{1}{2M_{Y}}\left[\Sigma_{Y}^{sH}-\Sigma_{Y}^{0H}\right]^{2},
  \label{eq:potential-depth}
\end{equation}
where $\Sigma_{Y}^{s(0)H}$ is the direct term of the baryon self-energy for the scalar (the time component of the vector) part. Then, we can determine the coupling constants, $g_{\sigma Y}$, following the values of potential depth around $n_0$ suggested from the experimental data of hypernuclei: $U_{\Lambda}^{(N)}=-28$ MeV, $U_{\Sigma}^{(N)} = +30$ MeV and $U_{\Xi}^{(N)} = -18$ MeV \citep{Schaffner:1993qj}.
The scalar strange coupling constants, $g_{\sigma^{\ast}Y}$, are restricted by the relation $U_{\Xi}^{(\Xi)}\simeq U_{\Lambda}^{(\Xi)}\simeq2U_{\Xi}^{(\Lambda)}\simeq2U_{\Lambda}^{(\Lambda)}$ \citep{Schaffner:1993qj,Schaffner:1995th,Yang:2008am}.
We here take $U_{\Lambda}^{(\Lambda)}\simeq-5$ MeV which has been implied by the Nagara event \citep{Takahashi:2001nm}.

Furthermore, the coupling constants for the vector mesons to hyperons, $g_{\omega Y}$, $g_{\phi Y}$, and $g_{\rho Y}$, are given by the following SU(3) relations:
\begin{align}
  g_{\omega\Lambda} = g_{\omega\Sigma}
	&= \frac{1}{1+\sqrt{3}z\tan\theta_{v}} g_{\omega N},
	\hspace{0.5cm}
	g_{\omega\Xi}
	= \frac{1-\sqrt{3}z\tan\theta_{v}}{1+\sqrt{3}z\tan\theta_{v}} g_{\omega N},
	\label{eq:SU3-omega} \\
	g_{\phi\Lambda} = g_{\phi\Sigma}
	&= \frac{-\tan\theta_{v}}{1+\sqrt{3}z\tan\theta_{v}} g_{\omega N},
	\hspace{0.5cm}
	g_{\phi\Xi}
	= -\frac{\sqrt{3}z+\tan\theta_{v}}{1+\sqrt{3}z\tan\theta_{v}} g_{\omega N},
	\label{eq:SU3-phi} \\
	g_{\rho N}
	&= \frac{1}{2} g_{\rho\Sigma} = g_{\rho\Xi},
	\hspace{0.5cm} g_{\rho\Lambda} = 0,
	\label{eq:SU6-rho}
\end{align}
with the values of $\theta_{v}$ and $z$ given below Equation \eqref{eq:SU3-mix} \citep{Miyatsu:2013yta,Miyatsu:2014wca,Rijken:2010zzb}.
Through the above relations, those couplings are determined once the values of $(g_{\omega N}, g_{\rho N})$ are given.
For the tensor couplings, we may again use the ratios, $f_{\omega Y}/g_{\omega Y}$, $f_{\phi Y}/g_{\phi Y}$, and $f_{\rho Y}/g_{\rho Y}$, suggested by the ESC08 model.
Furthermore, the coupling of $f_{\pi Y}$ is also chosen to be the value in the ESC08 model.

In the present calculations, we consider two parameter sets of the coupling constants for hyperons: one is the original set suggested by the ESC08 model (hereafter we call it ESC08Y), and the other is the values based on {\it naive} SU(3) symmetry (SU(3)Y).
In both cases, we use the coupling values, $g_{\sigma N}$, $g_{\omega N}$, $g_{\phi N}$ and $g_{\rho N}$, for SU(3) symmetry in Table \ref{tab:CCN}.  Then, in SU(3)Y, using Equations \eqref{eq:SU3-omega}--\eqref{eq:SU6-rho} and the values of $(g_{\omega N}, g_{\rho N})$ in Table \ref{tab:CCN}, the coupling constants for hyperons are calculated.  In contrast, in ESC08Y, we adopt the values of $(g_{\omega N}, g_{\rho N})$ presented in Table IV of \citet{Rijken:2010zzb} in the calculation of coupling constants for hyperons.
The couplings related to the hyperons are listed in Table \ref{tab:CCY}.

%%%%%%%%%%%%%%%%%%%%%%%%%%%%%%%%%%%%%%%%%%%%%%%%%%%%%%%%%%%%%%%%%%%%%
\subsection{Neutron star properties}
\label{subsec:result-NS}
%%%%%%%%%%%%%%%%%%%%%%%%%%%%%%%%%%%%%%%%%%%%%%%%%%%%%%%%%%%%%%%%%%%%%

The properties of neutron stars are, in general, estimated by solving the the Tolman-Oppenheimer-Volkoff (TOV) equation \citep{Tolman:1939jz,Oppenheimer:1939ne}.
Since the radius of a neutron star is remarkably sensitive to the EoS at very low densities, we use the EoS for nonuniform matter below $n_{B}=0.068$ fm$^{-3}$, where nuclei are taken into account using the Thomas-Fermi calculation \citep{Miyatsu:2013hea}.

\begin{figure}
\epsscale{.80}
\plotone{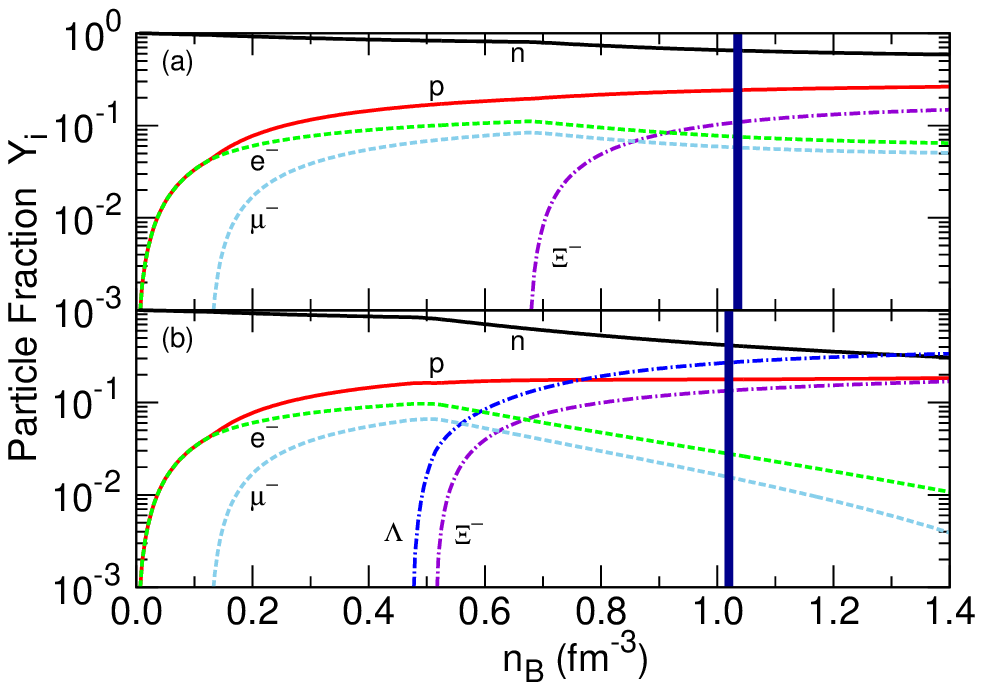}
\caption{\label{fig:Composition-hadron}
Particle fractions, $Y_{i}$, for hadronic matter in ESC08Y and SU(3)Y, which 
is defined as $Y_{i}=\rho_{i}/n_{B}$ with $\rho_{i}$ being the number densities of particle species $i=B,\ell$.
The upper panel (a) is for the case of ESC08Y, and the lower panel (b) is for the case of SU(3)Y.
The thick line shows the density at which a neutron star reaches the maximum-mass point by solving the TOV equation.
}
\end{figure}

In Figure \ref{fig:Composition-hadron}, we illustrate the particle fractions for hadronic matter in ESC08Y and SU(3)Y.
Although all octet baryons are considered in the present calculation, only $\Xi^{-}$ appears at 0.675 fm$^{-3}$ in ESC08Y, whereas, in SU(3)Y, the $\Lambda$ firstly appears at 0.475 fm$^{-3}$, followed by $\Xi^{-}$ at 0.515 fm$^{-3}$.
We also note that the Fock contribution suppresses the appearance of hyperons as compared with the RH calculation including the $\sigma^{\ast}$ and $\phi$ in SU(3) flavor symmetry (see Figure 1 in \citet{Miyatsu:2013yta}).
Due to the effect of the strange mesons, the threshold density of the $\Xi^{-}$ production in ESC08Y is higher than that in our previous calculation (see Figure 6 in \citet{Miyatsu:2013hea}).
Furthermore, the densities at which a neutron star reaches the maximum mass are almost the same in both cases.

\begin{figure}
\epsscale{.80}
\plotone{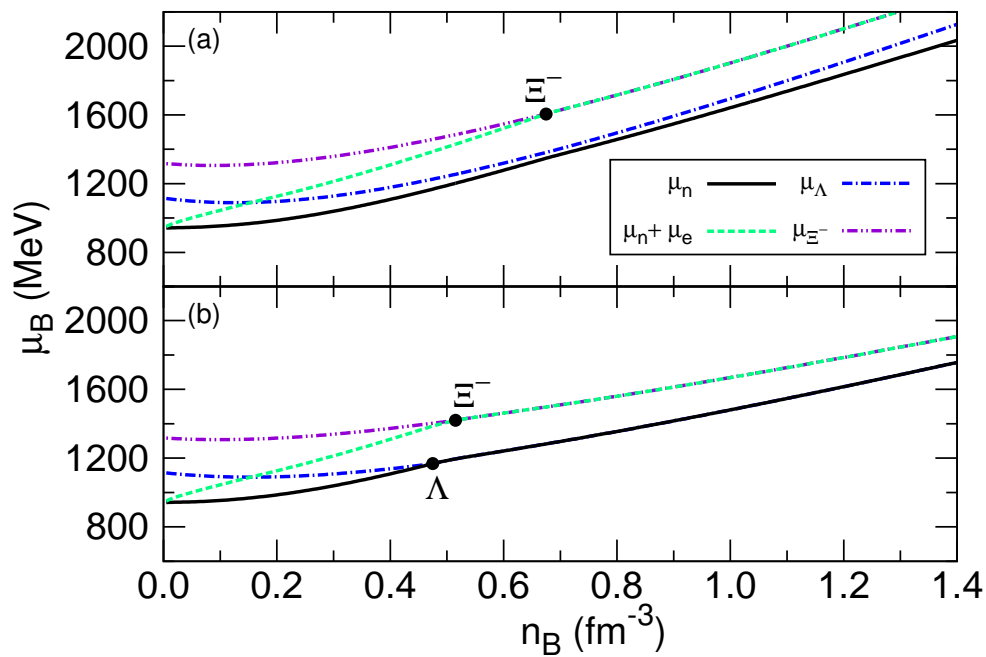}
\caption{\label{fig:Chemical-potential-hadron}
Chemical potentials for the $n$, $\Lambda$, $\Xi^{-}$ and $e^{-}$ in ESC08Y and SU(3)Y.
The filled circle denotes the onset of $\Lambda$ or $\Xi^{-}$.
The labels (a) and (b) are the same as in Figure \ref{fig:Composition-hadron}.
}
\end{figure}

The chemical potentials for the $n$, $\Lambda$, $\Xi^{-}$ and $e^{-}$ in hadronic matter are presented in Figure \ref{fig:Chemical-potential-hadron}.
When the $\beta$ equilibrium conditions given in Equations \eqref{eq:beta-neutral}--\eqref{eq:beta-plus} are satisfied, hyperons can be generated.
At low densities, the chemical potentials for the $\Lambda$ and $\Xi^{-}$ behaves similarly in both cases, because, as explained in Section \ref{subsec:result-matter}, the hyperon coupling constants are determined so as to reproduce the observed, potential depths at $n_{0}$.
However, their behaviors in ESC08Y show some difference from those in SU(3)Y above $n_{B} \simeq 0.4$ fm$^{-3}$.
In particular, the chemical potential for the $\Lambda$ in ESC08Y is slightly higher than that in SU(3)Y at middle and high densities.
As shown in Figure \ref{fig:Composition-hadron}, in the density region below 1.4 fm$^{-3}$, the $\Lambda$ disappears in ESC08Y, while it appears at 0.475 fm$^{-3}$ in SU(3)Y.
This is mainly because the repulsive force due to the $\omega$ and $\phi$ mesons strongly affects the chemical potential for the hyperons through the self-energy of the time component given in Equation \eqref{eq:BSE-time}.
As already mentioned in Table \ref{tab:CCY}, the coupling constant $g_{\omega Y}$ in ESC08Y is larger than that in SU(3)Y, and thus the absolute value of the hyperon self-energy of the time component becomes larger in ESC08Y as the density increases, especially at high densities.  It leads to the larger $\mu_{\Lambda}$.
Then, the hyperon creation in ESC08Y is suppressed compared with that in SU(3)Y.
In addition, the $\Sigma$ hyperon does not appear in both cases, because the $\Sigma$-hyperon potential in nuclear matter around $n_0$, $U_{\Sigma}^{(N)}$, is chosen to be repulsive.
Furthermore, it is found that the chemical potential for the nucleon at high densities affects the onset of the hyperon productions, and the difference between the chemical potentials in ESC08Y and SU(3)Y becomes about 160 MeV around 1.0 fm$^{-3}$, around which a neutron-star mass reaches the maximum value.
We note that, compared with the usual RH results, the Fock contribution also affects the chemical potential of neutron at high densities and hinders the hyperon creation \citep{Miyatsu:2011bc,Katayama:2012ge,Whittenbury:2013wma}.

\begin{figure}
\epsscale{.80}
\plotone{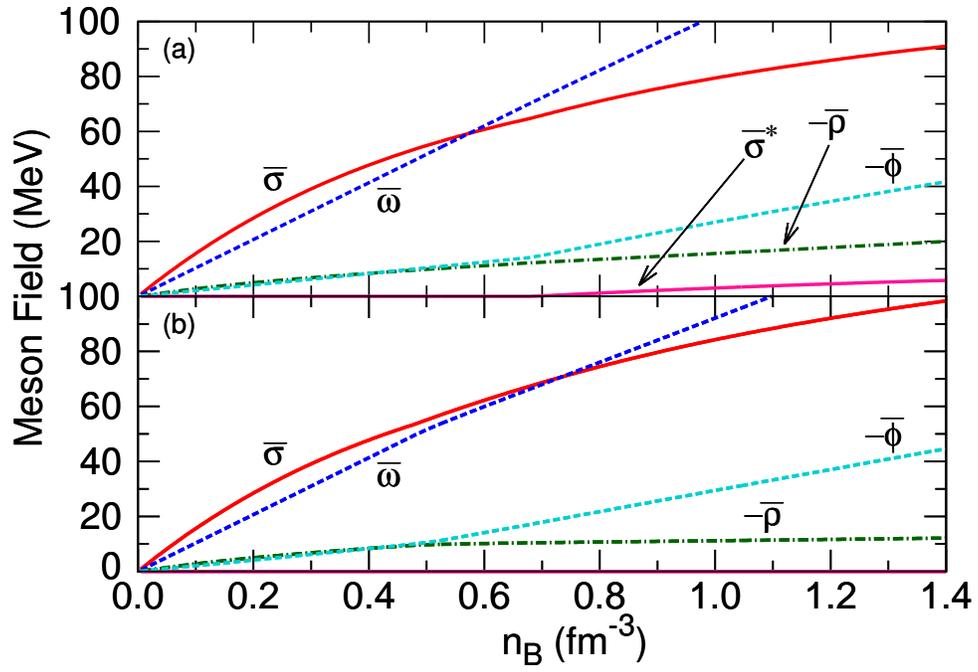}
\caption{\label{fig:Meson-field}
Meson fields in ESC08Y and SU(3)Y.
The labels (a) and (b) are the same as in Figure \ref{fig:Composition-hadron}.
}
\end{figure}

The meson fields are presented in Figure \ref{fig:Meson-field}.
The $\phi$ meson considerably contributes to the baryon interactions even at low densities because of the mixing effect in SU(3) flavor symmetry.
In ESC08Y, the $\sigma^{\ast}$ meson emerges above the density at which the $\Xi^{-}$ hyperon is created, because we assume that $g_{\sigma^{\ast}N}=0$ in the present calculation.
In contrast, the $\sigma^{\ast}$ meson is not introduced in SU(3)Y, because the reasonable properties of hypernuclei can be reproduced even if the attractive force due to the $\sigma^{\ast}$ is not included \citep{Schaffner:1993qj,Schaffner:1995th,Takahashi:2001nm,Yang:2008am}.

\begin{figure}
\epsscale{.80}
\plotone{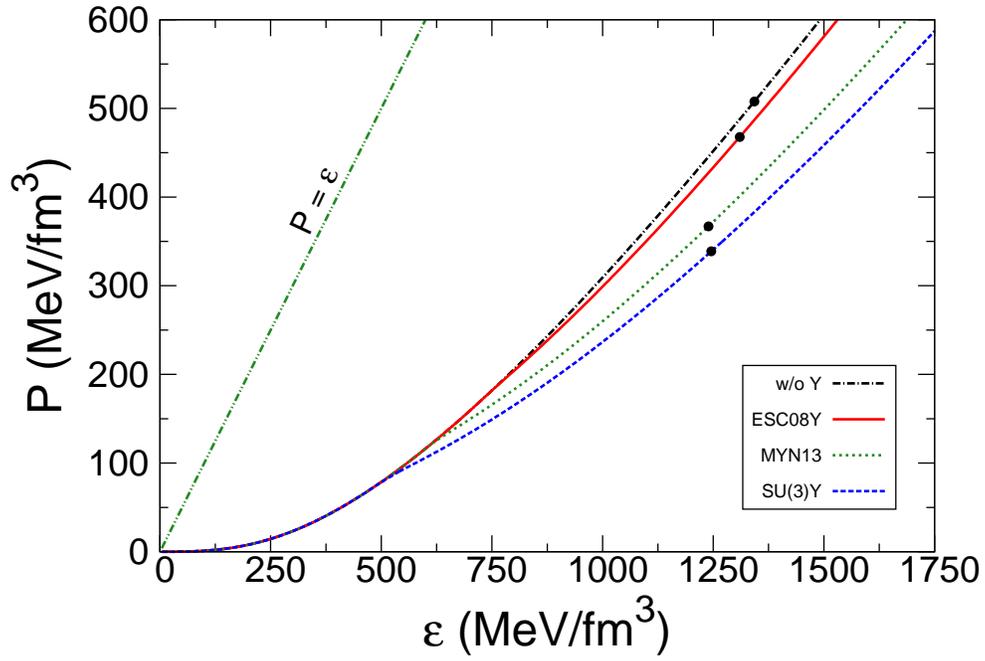}
\caption{\label{fig:EoS-hadron}
Equations of state for hadronic matter in ESC08Y and SU(3)Y.
We also present our previous EoS denoted by MYN13 \citep{Miyatsu:2013hea}.
The filled circle shows the point at which a neutron star reaches the maximum mass.
}
\end{figure}

\begin{figure}
\epsscale{.80}
\plotone{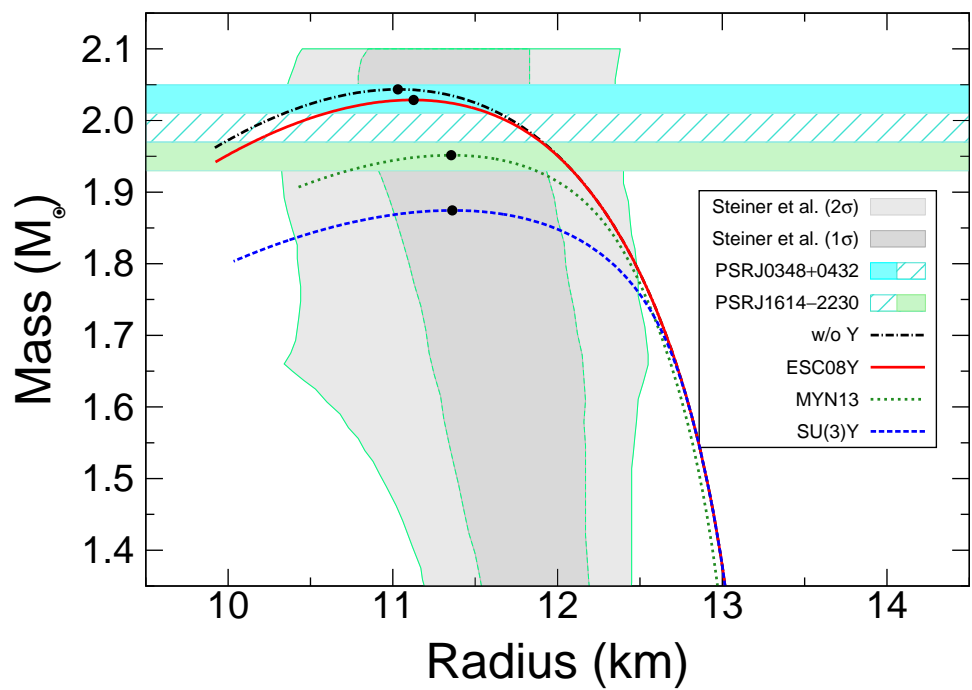}
\caption{\label{fig:TOV-hadron}
Mass-radius relations for hadronic stars in ESC08Y and SU(3)Y.
The filled circle shows the maximum-mass point of a neutron star.
The shaded bands show the recent $2M_{\odot}$ constraint from the measurements of PSR J1614-2230 and PSR J0348+0432 \citep{Demorest:2010bx,Antoniadis:2013pzd}.
The two shaded areas show the $1\sigma$ and $2\sigma$ confidence limits for neutron-star radii of fixed mass obtained from the analysis by \citet{Steiner:2010fz}.
}
\end{figure}

In Figures \ref{fig:EoS-hadron} and \ref{fig:TOV-hadron}, we respectively show the EoS and the mass of a neutron star as a function of the neutron-star radius in ESC08Y and  SU(3)Y.
As well known, the inclusion of hyperons makes the EoS soft, and thus the maximum mass of a neutron star is reduced (see the curves for SU(3)Y in Figures \ref{fig:EoS-hadron} and \ref{fig:TOV-hadron}).
However, due to the suppression of the hyperon appearance and the large chemical potential for neutrons at high density, the EoS in ESC08Y keeps stiffness and the maximum mass of a neutron star can satisfy the recent $2M_{\odot}$ constraint from the measurements of PSR J1614-2230 and PSR J0348+0432 \citep{Demorest:2010bx,Antoniadis:2013pzd}.
Compared with our previous calculation denoted by MYN13 \citep{Miyatsu:2013hea}, the maximum mass in ESC08Y becomes heavier from $1.951M_{\odot}$ to $2.029M_{\odot}$.
This is mainly because of the additional, repulsive force due to the strange, vector $\phi$ meson.
This fact can also be seen in the result based on the RH calculation in SU(3) flavor symmetry \citep{Miyatsu:2013yta,Miyatsu:2014wca,Weissenborn:2011ut,Lopes:2013cpa}.
In contrast, the maximum mass in SU(3)Y is below the $2M_{\odot}$ constraint from the pulsar observations.
The radius of a neutron star around the maximum mass lies in the shaded areas obtained from theoretical analysis using  the Markov chain Monte Carlo within Bayesian framework \citep{Steiner:2010fz}, but the radius of the neutron star with  $1.4M_{\odot}$ in the present calculations is about 1 km larger than the range predicted by Bayesian analysis.

%%%%%%%%%%%%%%%%%%%%%%%%%%%%%%%%%%%%%%%%%%%%%%%%%%%%%%%%%%%%%%%%%%%%%
\subsection{Quark coexistence in neutron stars}
\label{subsec:result-HS}
%%%%%%%%%%%%%%%%%%%%%%%%%%%%%%%%%%%%%%%%%%%%%%%%%%%%%%%%%%%%%%%%%%%%%

As shown in Figure \ref{fig:Chemical-potential-hadron}, the chemical potentials for baryons increase as the density grows.
Therefore, quarks may also be generated in the core region if the $\beta$ equilibrium conditions given in Equations \eqref{eq:beta-neutral}--\eqref{eq:beta-lepton} are satisfied at some critical baryon density, $n_{B}^{(c)}$.
In the MP where quarks coexist with hyperons as well as nucleons and leptons, we impose the conditions of equal pressure and {\it global} charge neutrality (see Equations \eqref{eq:Gibbs-criterion-press} and \eqref{eq:Gibbs-criterion-charge}) in the range of the volume fraction, $0<\chi<1$.

\begin{figure}
\epsscale{.80}
\plotone{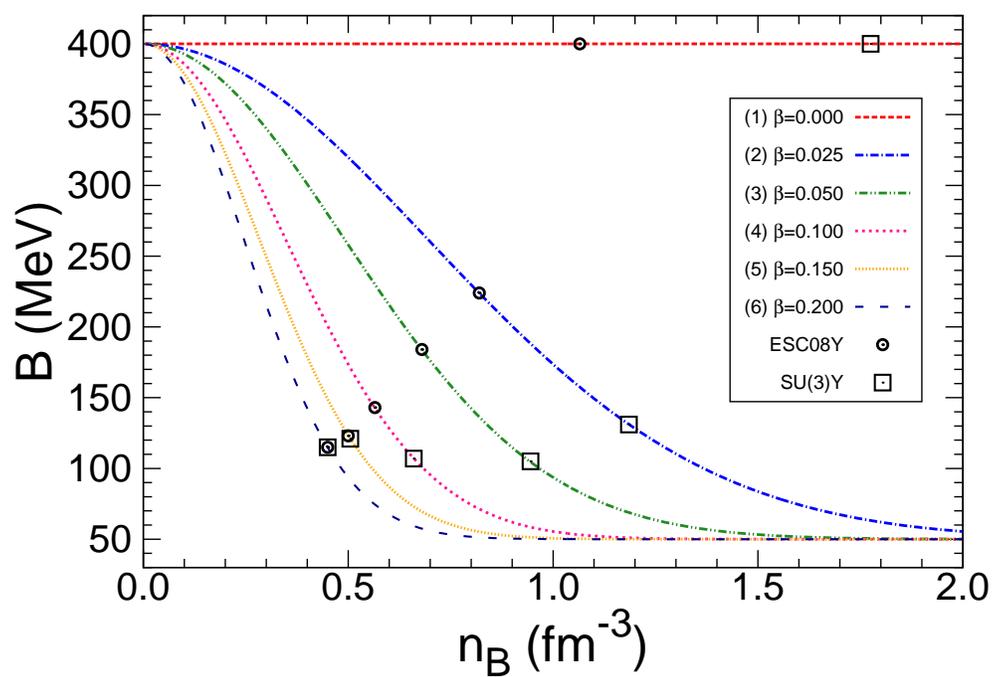}
\caption{\label{fig:DDBagc}
Density-dependence of bag constant given in Equation \eqref{eq:DDbag}.
The circles (squares) show the critical points in ESC08Y (SU(3)Y) (see also Table \ref{tab:Phase-transition-properties}).
}
\end{figure}

In the present calculations, the phase transition from hadrons to quarks can be controlled by the density-dependent bag constant in Equation \eqref{eq:DDbag}.
In order to study the influence of the phase transition on the properties of neutron stars with hadrons, leptons, and quarks (i.e. hybrid stars), as shown in Table \ref{tab:Phase-transition-properties}, we examine six cases, where the parameter $\beta$ varies between 0 to 0.2, in ESC08Y and SU(3)Y.
In Figure \ref{fig:DDBagc}, the density-dependent bag constant is presented as a function of the total baryon density.
It is found that the critical bag constant, $B^{(c)}$, and the critical chemical potential, $\mu_{B}^{(c)}$, become smaller as the parameter $\beta$ is larger.

\begin{figure}
\epsscale{1.00}
\plotone{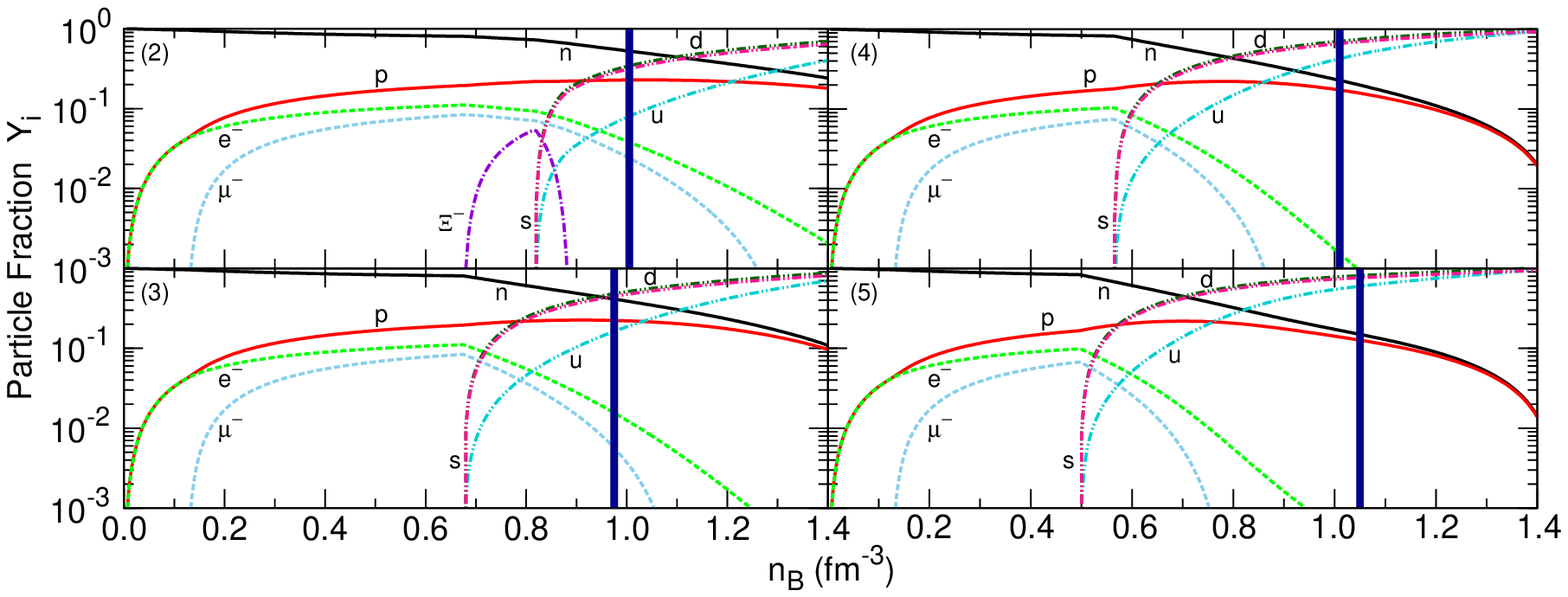}
\caption{\label{fig:Composition-hybrid-ESC08Y}
Particle fractions, $Y_{i}$, for hybrid-star matter in ESC08Y.
The labels (2)--(5) designate the case number in Table \ref{tab:Phase-transition-properties}.
}
\end{figure}

\begin{figure}
\epsscale{1.00}
\plotone{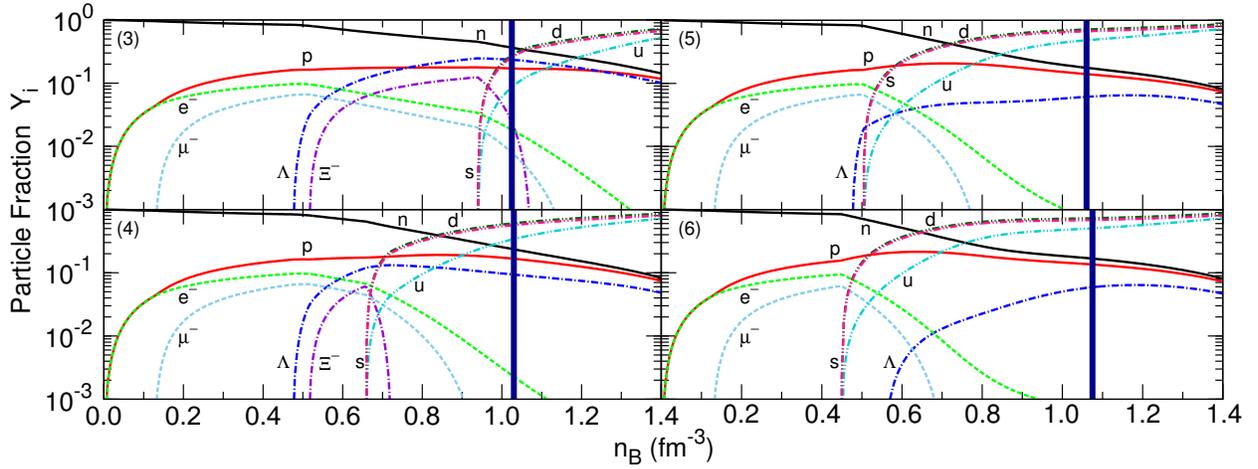}
\caption{\label{fig:Composition-hybrid-SU3Y}
Particle fractions, $Y_{i}$, for hybrid-star matter in SU(3)Y.
The labels (3)--(6) designate the case number in Table \ref{tab:Phase-transition-properties}.
}
\end{figure}

Relevant particle fractions for hybrid-star matter are given in Figures \ref{fig:Composition-hybrid-ESC08Y} and \ref{fig:Composition-hybrid-SU3Y}.
With the increase of the parameter $\beta$ in the density-dependent bag constant, the threshold densities of quark productions move toward lower densities, and the hyperon population at high densities is suppressed in both hadronic cases, ESC08Y and SU(3)Y.
Due to the quark productions, the $\Xi^{-}$ completely disappears at the densities below 1.4 fm$^{-3}$, while, only in SU(3)Y, the $\Lambda$ can still be generated even in the MP because of its neutral charge (see the panel (6) in Figure \ref{fig:Composition-hybrid-SU3Y}).
In addition, the lepton fractions dwindle immediately at the densities above the onset of the quarks, because negatively charged particles are replaced by an abundance of quarks to conserve the charge neutrality.

\begin{figure}
\epsscale{.80}
\plotone{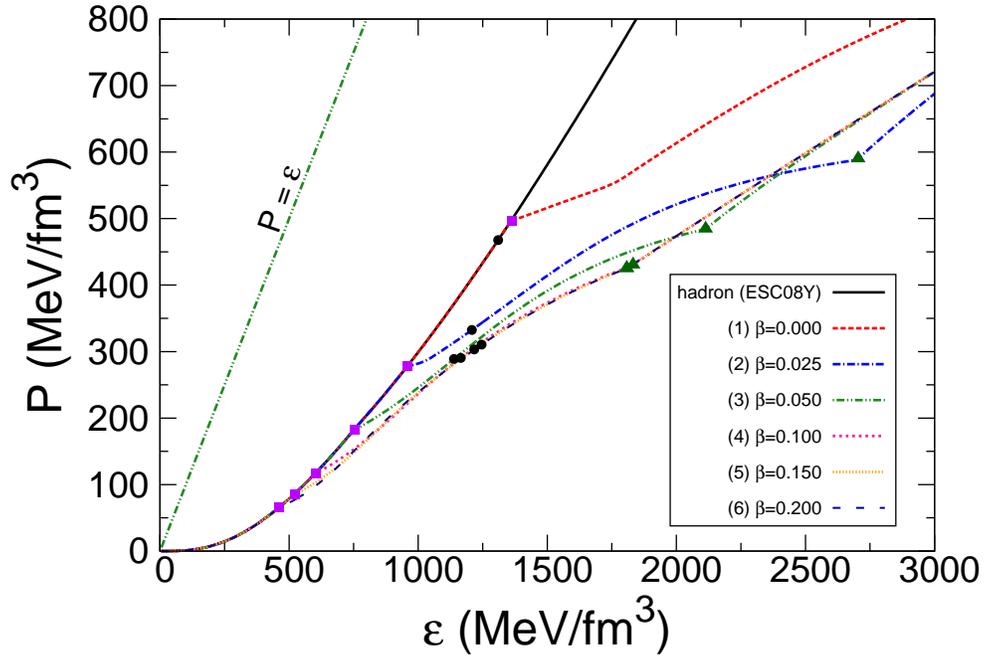}
\caption{\label{fig:EoS-hybrid-ESC08Y}
Equations of state for hybrid-star matter in ESC08Y.
The filled circles show the points at the maximum mass of a neutron star, and the squares (triangles) mean the beginning (end) points of the mixed phase, i.e. $\chi=0$ and $\chi=1$ in Equations \eqref{eq:total-engy-density-mixed-phase} and \eqref{eq:baryon-number-density-mixed-phase}.
}
\end{figure}

In Figure \ref{fig:EoS-hybrid-ESC08Y}, we present the EoS for hybrid-star matter in ESC08Y.
In general, the quark productions soften the EoS, as well known.
Increasing the parameter $\beta$, the EoS becomes softer because the threshold densities of the quarks move toward lower densities, as seen in Table \ref{tab:Phase-transition-properties} and Figure \ref{fig:Composition-hybrid-ESC08Y}.
Furthermore, the pure QP emerges at lower energy densities as the parameter $\beta$ increases.

\begin{figure}
\epsscale{.80}
\plotone{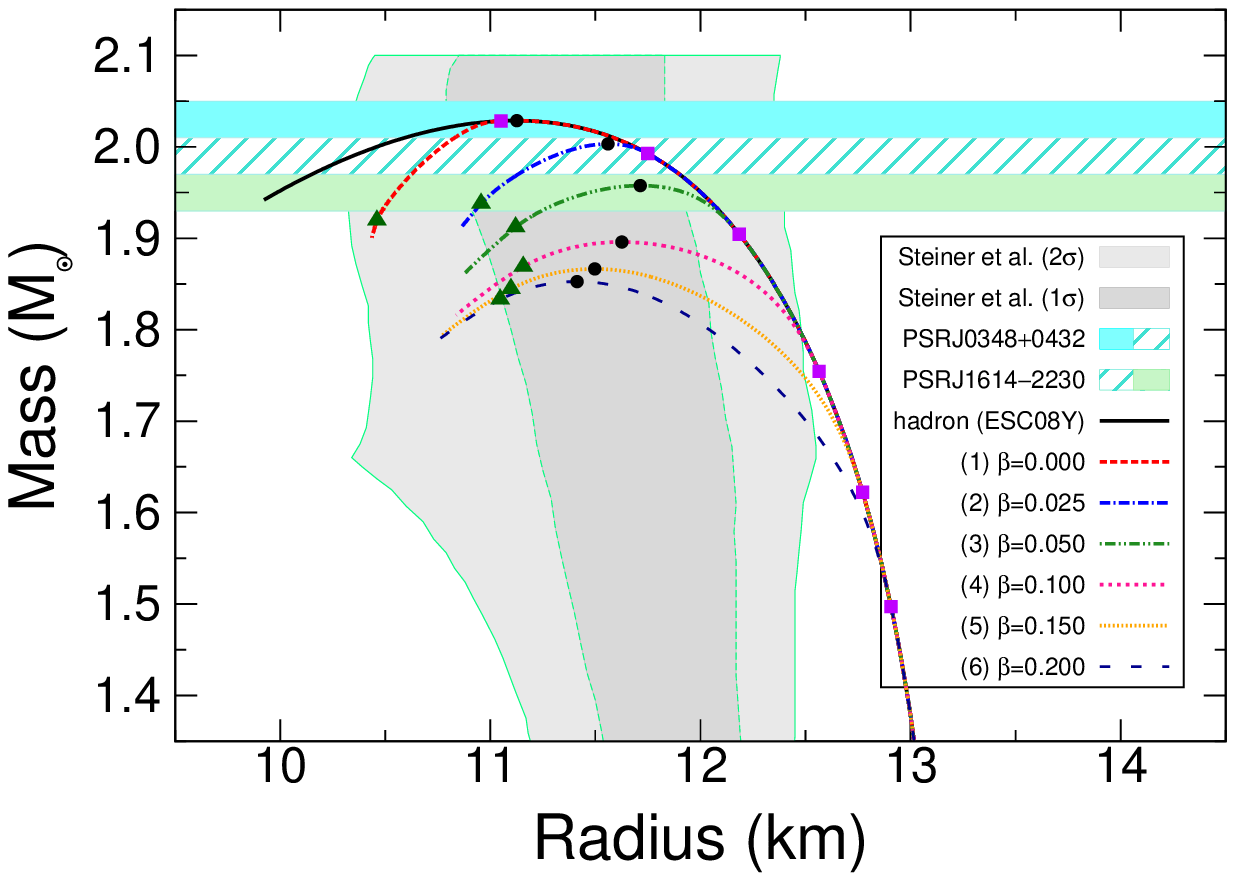}
\caption{\label{fig:TOV-hybrid-ESC08Y}
Mass-radius relations for hybrid stars in ESC08Y.
The shaded bands and areas are the same as in Figure \ref{fig:TOV-hadron}.
The filled circles, squares, and triangles are explained in the caption of Figure \ref{fig:EoS-hybrid-ESC08Y}.
}
\end{figure}

We summarize the properties of neutron stars in Table \ref{tab:NS-properties}, and illustrate the mass-radius relation of hybrid stars in Figure \ref{fig:TOV-hybrid-ESC08Y}.
It is found that, except the case (1), the maximum mass of a neutron star is reduced because the quark productions make the EoS soft.
Moreover, the maximum mass is realized in the MP, and the transition to pure quark matter occurs only in the neutron star which already lies on the gravitationally unstable branch of the stellar sequence.
Meanwhile, in the case (1), the threshold density of the quark productions, $n_{B}^{(c)}$, follows the central density, $n_{c}$, at the maximum-mass point, and then the maximum mass appears before the QP transition.
In the case (2) of ESC08Y, the maximum mass reaches $2.003M_{\odot}$ even if quark matter appears in the core of a neutron star.
Therefore, the present EoS for hybrid-star matter can reasonably explain the recent mass constraint from astrophysical observations even if hyperons and quarks are included.
As the properties of hybrid stars in SU(3)Y are similar to those in ESC08Y, we here do not present them.
However, in SU(3)Y including quark degrees of freedom it is impossible to support the massive neutron stars (see Table \ref{tab:NS-properties}).

\begin{figure}
\epsscale{.80}
\plotone{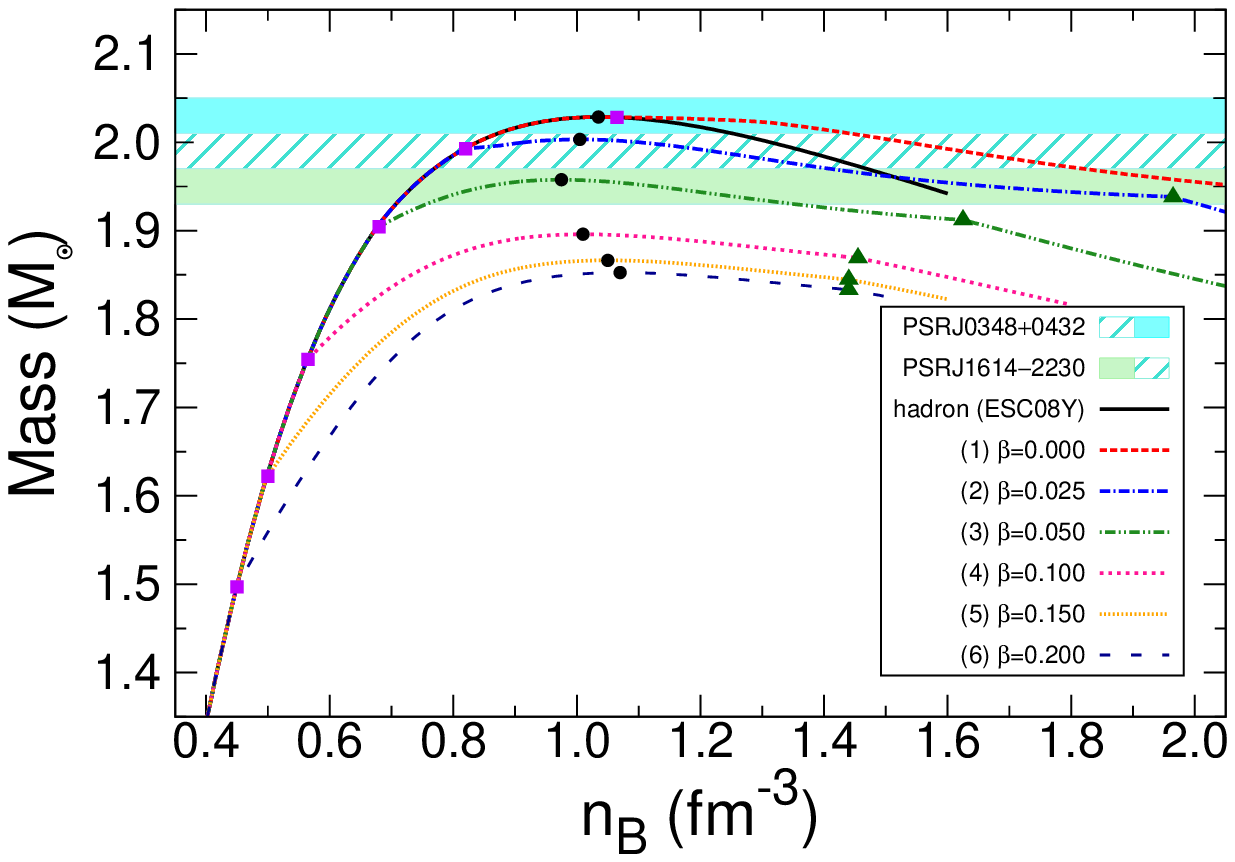}
\caption{\label{fig:Mass-density-hybrid-ESC08Y}
Mass-density relations for hybrid stars in ESC08Y.
The shaded bands and areas are the same as in Figure \ref{fig:TOV-hadron}.
The filled circles, squares, and triangles are explained in the caption of Figure \ref{fig:EoS-hybrid-ESC08Y}.
}
\end{figure}

The neutron-star mass as a function of the total baryon density in ESC08Y is presented in Figure \ref{fig:Mass-density-hybrid-ESC08Y}.
We find that, in spite of varying the parameter $\beta$ in the density-dependent bag constant, all the maximum masses exist around the density of 1.0 fm$^{-3}$.
In contrast, the threshold densities of the quark productions are extremely sensitive to the parameter $\beta$, as already seen in Figures \ref{fig:Composition-hybrid-ESC08Y} and \ref{fig:Composition-hybrid-SU3Y}, and the end points of the MP  also strongly depend on the parameter $\beta$.
Even if the quark degrees of freedom as well as hadrons and leptons are considered in the core of a neutron star, the EoS with the parameter range, $\beta\leq0.025$, is consistent with the observation of heavy ($\sim 2M_{\odot}$) neutron stars.
From Table \ref{tab:Phase-transition-properties}, we can finally find the lower limit of the critical chemical potential for $\beta=0.025$, $\mu_{B}^{(c)}\sim1.5$ GeV, which is almost the same value as in the calculation using the Polyakov-loop extended Nambu-Jona-Lasinio model with the vector-type four-body interaction and the hadron resonance gas model with the volume-exclusion effect \citep{Sasaki:2013mha}.

%%%%%%%%%%%%%%%%%%%%%%%%%%%%%%%%%%%%%%%%%%%%%%%%%%%%%%%%%%%%%%%%%%%%%%%%%%%%%%%%%%%%%%%%%%%%%%%%%%%
\section{Summary}
\label{sec:Summary}
%%%%%%%%%%%%%%%%%%%%%%%%%%%%%%%%%%%%%%%%%%%%%%%%%%%%%%%%%%%%%%%%%%%%%%%%%%%%%%%%%%%%%%%%%%%%%%%%%%%

We have constructed the EoS for neutron stars with hyperons and quarks, and studied the properties of neutron stars with and without deconfined quarks in the core region.
Under RHF approximation, the EoS for hadronic matter has been calculated by including of the strange ($\sigma^{\ast}$ and $\phi$) mesons as well as the light non-strange ($\sigma$, $\omega$, $\bm{\pi}$, and $\bm{\rho}$) mesons.
In addition, we have used the chiral quark-meson coupling (CQMC) model to take the variation of the in-medium baryon structure into account \citep{Nagai:2008ai,Miyatsu:2010zz,Saito:2010zw}.
The EoS for quark matter has been calculated using the MIT bag model with one-gluon-exchange interaction, and the phase transition from hadronic matter to quark matter has been achieved under Gibbs criteria for chemical equilibrium \citep{Glendenning:1992vb,Glendenning:2001pe}.

In the present calculations for hadronic EoS, we have examined the extension from SU(6) spin-flavor symmetry based on the quark model to SU(3) flavor symmetry in determining the isoscalar, vector-meson couplings to the octet baryons within RHF approximation \citep{Miyatsu:2013yta,Miyatsu:2014wca,Weissenborn:2011ut,Lopes:2013cpa}.
We have found that, in both cases of the SU(6) and the SU(3) symmetries, the calculated properties of nuclear matter at $n_{0}$, for instance, the incompressibility, the slope parameter of the symmetry energy, etc. are consistent with the experimental data and/or values predicted by other theoretical calculations.

For hyperon coupling constants, we have adopted two parameter sets: one is the original set suggested by the ESC08 model (ESC08Y), and the other is based on {\it naive} SU(3) symmetry (SU(3)Y).
We have found that, compared with our previous results \citep{Miyatsu:2013hea}, the maximum mass in ESC08Y becomes heavier from $1.95M_{\odot}$ to $2.03M_{\odot}$,
because of the additional, repulsive force due to the strange vector ($\phi$) meson and the hyperon suppression due to the large chemical potential of neutrons caused by the Fock contribution.
While, in SU(3)Y, it is difficult to satisfy the $2M_{\odot}$ constraint from the recent pulsar observations, because the hyperon appearance move toward lower densities, due to the small chemical potentials of the $\Lambda$ and $\Xi^{-}$, and it  consequently makes the EoS softer.

In order to study the effect of the phase transition on the properties of hybrid stars, we have introduced the density-dependent bag constant for quark matter \citep{Burgio:2001mk,Burgio:2002sn}.
If we presume the first order phase transition under Gibbs criteria \citep{Glendenning:1992vb,Glendenning:2001pe}, quark matter suppresses the population of hadrons and leptons, and it especially hinders the hyperon production.
We have also found that the maximum mass in the case (2) of ESC08Y can reaches $2.003M_{\odot}$ (see Table \ref{tab:NS-properties}) even when hyperon and quark degrees of freedom are taken into account.
We note that, even if quarks interact {\it without} strong, repulsive vector interaction, the present result can satisfy the mass constraint from the recent astrophysical observations \citep{Demorest:2010bx,Antoniadis:2013pzd}.
Then, we have determined the lower bounds of the critical density for the phase transition, $n_{B}^{(c)}\sim0.82$ fm$^{-3}$, and the critical chemical potential, $\mu_{B}^{(c)}\sim1.5$ GeV, which are consistent with the recent calculation by \citet{Sasaki:2013mha}.
In addition, our results have shown that the transition to pure quark matter occurs only in the neutron star lying on the gravitationally unstable branch of the stellar sequence.

Finally, we give several comments on the future work.
It may be important to consider the mixing effect of $\Lambda\Sigma$ channel in the Fock diagram through the $\bm{\pi}$ or $\bm{\rho}$ meson exchange.
Furthermore, the $\bm{K}$ or $\bm{K}^{\ast}$ meson exchange between two baryons may be desirable in order to take the mixing of $N$-$\Lambda$, $\Lambda$-$\Xi$, $N$-$\Sigma$, or $\Lambda$-$\Sigma$ into account, because these mixing may affect the EoS for hadronic matter, and consequently the particle fractions in the core region may be changed \citep{Katayama:2015dga,Yamamoto:2014jga}.
For the possibility of other exotic degrees of freedom, it is also interesting to include the effect of meson condensates  \citep{Ryu:2007qz,Muto:2008wg} in the RHF calculation for neutron stars.

%%%%%%%%%%%%%%%%%%%%%%%%%%%%%%%%%%%%%%%%%%%%%%%%%%%%%%%%%%%%%%%%%%%%%%%%%%%%%%%%%%%%%%%%%%%%%%%%%%%
\acknowledgments
This work was supported by the National Research Foundation of Korea (Grant No. NRF-2014R1A2A2A05003548).
%%%%%%%%%%%%%%%%%%%%%%%%%%%%%%%%%%%%%%%%%%%%%%%%%%%%%%%%%%%%%%%%%%%%%%%%%%%%%%%%%%%%%%%%%%%%%%%%%%%

\clearpage

%%%%%%%%%%%%%%%%%%%%%%%%%%%%%%%%%%%%%%%%%%%%%%%%%%%%%%%%%%%%%%%%%%%%%%%%%%%%%%%
%%%%%%%%%%%%%%%%%%%%%%%%%%%%%%%%%%%%%%%%%%%%%%%%%%%%%%%%%%%%%%%%%%%%%%%%%%%%%%%

%%%%%%%%%%%%%%%%%%%%%%%%%%%%%%%%%%%%%%%%%%%%%%%%%%%%%%%%%%%%%%%%%%%%%%%%%%%%%%%
%%%%%%%%%%%%%%%%%%%%%%%%%%%%%%%%%%%%%%%%%%%%%%%%%%%%%%%%%%%%%%%%%%%%%%%%%%%%%%%

%%%%%%%%%%%%%%%%%%%%%%%%%%%%%%%%%%%%%%%%%%%%%%%%%%%%%%%%%%%%%%%%%%%%%%%%%%%%%%%
%%%%%%%%%%%%%%%%%%%%%%%%%%%%%%%%%%%%%%%%%%%%%%%%%%%%%%%%%%%%%%%%%%%%%%%%%%%%%%%

%%%%%%%%%%%%%%%%%%%%%%%%%%%%%%%%%%%%%%%%%%%%%%%%%%%%%%%%%%%%%%%%%%%%%%%%%%%%%%%
%%%%%%%%%%%%%%%%%%%%%%%%%%%%%%%%%%%%%%%%%%%%%%%%%%%%%%%%%%%%%%%%%%%%%%%%%%%%%%%

%%%%%%%%%%%%%%%%%%%%%%%%%%%%%%%%%%%%%%%%%%%%%%%%%%%%%%%%%%%%%%%%%%%%%%%%%%%%%%%
%%%%%%%%%%%%%%%%%%%%%%%%%%%%%%%%%%%%%%%%%%%%%%%%%%%%%%%%%%%%%%%%%%%%%%%%%%%%%%%

%%%%%%%%%%%%%%%%%%%%%%%%%%%%%%%%%%%%%%%%%%%%%%%%%%%%%%%%%%%%%%%%%%%%%%%%%%%%%%%
%%%%%%%%%%%%%%%%%%%%%%%%%%%%%%%%%%%%%%%%%%%%%%%%%%%%%%%%%%%%%%%%%%%%%%%%%%%%%%%

\clearpage

%%%%%%%%%%%%%%%%%%%%%%%%%%%%%%%%%%%%%%%%%%%%%%%%%%%%%%%%%%%%%%%%%%%%%%%%%%%%%%%
%%%%%%%%%%%%%%%%%%%%%%%%%%%%%%%%%%%%%%%%%%%%%%%%%%%%%%%%%%%%%%%%%%%%%%%%%%%%%%%

%%%%%%%%%%%%%%%%%%%%%%%%%%%%%%%%%%%%%%%%%%%%%%%%%%%%%%%%%%%%%%%%%%%%%%%%%%%%%%%
%%%%%%%%%%%%%%%%%%%%%%%%%%%%%%%%%%%%%%%%%%%%%%%%%%%%%%%%%%%%%%%%%%%%%%%%%%%%%%%

%%%%%%%%%%%%%%%%%%%%%%%%%%%%%%%%%%%%%%%%%%%%%%%%%%%%%%%%%%%%%%%%%%%%%%%%%%%%%%%
%%%%%%%%%%%%%%%%%%%%%%%%%%%%%%%%%%%%%%%%%%%%%%%%%%%%%%%%%%%%%%%%%%%%%%%%%%%%%%%

%%%%%%%%%%%%%%%%%%%%%%%%%%%%%%%%%%%%%%%%%%%%%%%%%%%%%%%%%%%%%%%%%%%%%%%%%%%%%%%
%%%%%%%%%%%%%%%%%%%%%%%%%%%%%%%%%%%%%%%%%%%%%%%%%%%%%%%%%%%%%%%%%%%%%%%%%%%%%%%

%%%%%%%%%%%%%%%%%%%%%%%%%%%%%%%%%%%%%%%%%%%%%%%%%%%%%%%%%%%%%%%%%%%%%%%%%%%%%%%
%%%%%%%%%%%%%%%%%%%%%%%%%%%%%%%%%%%%%%%%%%%%%%%%%%%%%%%%%%%%%%%%%%%%%%%%%%%%%%%

\clearpage

%%%%%%%%%%%%%%%%%%%%%%%%%%%%%%%%%%%%%%%%%%%%%%%%%%%%%%%%%%%%%%%%%%%%%%%%%%%%%%%
\begin{deluxetable}{lcccc}
% \tabletypesize{\footnotesize}
\tablewidth{0pt}
\tablecaption{
Values of $a_{B}$, $b_{B}$, $a_{B}^{\prime}$, and $b_{B}^{\prime}$ for the octet baryons in the CQMC model.
}
\tablehead{
\colhead{$B$} & \colhead{$a_{B}$ (fm)} & \colhead{$b_{B}$} & \colhead{$a_{B}^{\prime}$ (fm)} & \colhead{$b_{B}^{\prime}$} }
\startdata
$N$           & 0.118                  & 1.04              & \nodata                         & \nodata                    \\
$\Lambda$     & 0.122                  & 1.09              & 0.290                           & 1.00                       \\
$\Sigma$      & 0.184                  & 1.02              & 0.277                           & 1.15                       \\
$\Xi$         & 0.181                  & 1.15              & 0.292                           & 1.04                       \\
\enddata
\label{tab:CQMC-parameter}
\tablecomments{
We assume that the scalar strange ($\sigma^{\ast}$) meson does not couple to the nucleon.
}
\end{deluxetable}
%%%%%%%%%%%%%%%%%%%%%%%%%%%%%%%%%%%%%%%%%%%%%%%%%%%%%%%%%%%%%%%%%%%%%%%%%%%%%%%

%%%%%%%%%%%%%%%%%%%%%%%%%%%%%%%%%%%%%%%%%%%%%%%%%%%%%%%%%%%%%%%%%%%%%%%%%%%%%%%
\begin{deluxetable}{lcccc}
% \rotate
\tabletypesize{\small}
\tablewidth{0pt}
\tablecaption{
Functions $A_{i}$, $B_{i}$, $C_{i}$, and $D_{i}$.
}
\tablehead{
\colhead{$i$} & \colhead{$A_{i}$} & \colhead{$B_{i}$} & \colhead{$C_{i}$} & \colhead{$D_{i}$} }
\startdata
$\sigma$ &
$g_{\sigma B}^{2}(\bar{\sigma})\Theta_{\sigma}$ &
$g_{\sigma B}^{2}(\bar{\sigma})\Theta_{\sigma}$ &
$-2g_{\sigma B}^{2}(\bar{\sigma})\Phi_{\sigma}$ &
\nodata \\
$\sigma^{\ast}$ &
$g_{\sigma^{\ast}B}^{2}(\bar{\sigma}^{\ast})\Theta_{\sigma^{\ast}}$ &
$g_{\sigma^{\ast}B}^{2}(\bar{\sigma}^{\ast})\Theta_{\sigma^{\ast}}$ &
$-2g_{\sigma^{\ast}B}^{2}(\bar{\sigma}^{\ast})\Phi_{\sigma^{\ast}}$ &
\nodata \\
$\omega_{VV}$ &
$2g_{\omega B}^{2}\Theta_{\omega}$ &
$-4g_{\omega B}^{2}\Theta_{\omega}$ &
$-4g_{\omega B}^{2}\Phi_{\omega}$ &
\nodata \\
$\omega_{TT}$ &
% $-\left(\frac{f_{\omega B}}{2\mathcal{M}}\right)^{2}m_{\omega}^{2}\Theta_{\omega}$ &
$-\left(f_{\omega B}/2\mathcal{M}\right)^{2}m_{\omega}^{2}\Theta_{\omega}$ &
% $-3\left(\frac{f_{\omega B}}{2\mathcal{M}}\right)^{2}m_{\omega}^{2}\Theta_{\omega}$ &
$-3\left(f_{\omega B}/2\mathcal{M}\right)^{2}m_{\omega}^{2}\Theta_{\omega}$ &
% $4\left(\frac{f_{\omega B}}{2\mathcal{M}}\right)^{2}\Psi_{\omega}$ &
$4\left(f_{\omega B}/2\mathcal{M}\right)^{2}\Psi_{\omega}$ &
\nodata \\
$\omega_{VT}$ &
\nodata &
\nodata &
\nodata &
% $12\frac{f_{\omega B}g_{\omega B}}{2\mathcal{M}}\Gamma_{\omega}$ \\
$12\left(f_{\omega B}g_{\omega B}/2\mathcal{M}\right)\Gamma_{\omega}$ \\
$\phi_{VV}$ &
$2g_{\phi B}^{2}\Theta_{\phi}$ &
$-4g_{\phi B}^{2}\Theta_{\phi}$ &
$-4g_{\phi B}^{2}\Phi_{\phi}$ &
\nodata \\
$\phi_{TT}$ &
% $-\left(\frac{f_{\phi B}}{2\mathcal{M}}\right)^{2}m_{\phi}^{2}\Theta_{\phi}$ &
$-\left(f_{\phi B}/2\mathcal{M}\right)^{2}m_{\phi}^{2}\Theta_{\phi}$ &
% $-3\left(\frac{f_{\phi B}}{2\mathcal{M}}\right)^{2}m_{\phi}^{2}\Theta_{\phi}$ &
$-3\left(f_{\phi B}/2\mathcal{M}\right)^{2}m_{\phi}^{2}\Theta_{\phi}$ &
% $4\left(\frac{f_{\phi B}}{2\mathcal{M}}\right)^{2}\Psi_{\phi}$ &
$4\left(f_{\phi B}/2\mathcal{M}\right)^{2}\Psi_{\phi}$ &
\nodata \\
$\phi_{VT}$ &
\nodata &
\nodata &
\nodata &
% $12\frac{f_{\phi B}g_{\phi B}}{2\mathcal{M}}\Gamma_{\phi}$ \\
$12\left(f_{\phi B}g_{\phi B}/2\mathcal{M}\right)\Gamma_{\phi}$ \\
$\rho_{VV}$ &
$2g_{\rho B}^{2}\Theta_{\rho}$ &
$-4g_{\rho B}^{2}\Theta_{\rho}$ &
$-4g_{\rho B}^{2}\Phi_{\rho}$ &
\nodata- \\
$\rho_{TT}$ &
% $-\left(\frac{f_{\rho B}}{2\mathcal{M}}\right)^{2}m_{\rho}^{2}\Theta_{\rho}$ &
$-\left(f_{\rho B}/2\mathcal{M}\right)^{2}m_{\rho}^{2}\Theta_{\rho}$ &
% $-3\left(\frac{f_{\rho B}}{2\mathcal{M}}\right)^{2}m_{\rho}^{2}\Theta_{\rho}$ &
$-3\left(f_{\rho B}/2\mathcal{M}\right)^{2}m_{\rho}^{2}\Theta_{\rho}$ &
% $4\left(\frac{f_{\rho B}}{2\mathcal{M}}\right)^{2}\Psi_{\rho}$ &
$4\left(f_{\rho B}/2\mathcal{M}\right)^{2}\Psi_{\rho}$ &
\nodata \\
$\rho_{VT}$ &
\nodata &
\nodata &
\nodata &
% $12\frac{f_{\rho B}g_{\rho B}}{2\mathcal{M}}\Gamma_{\rho}$ \\
$12\left(f_{\rho B}g_{\rho B}/2\mathcal{M}\right)\Gamma_{\rho}$ \\
$\pi_{pv}$ &
$-f_{\pi B}^{2}\Theta_{\pi}$ &
$-f_{\pi B}^{2}\Theta_{\pi}$ &
% $2\left(\frac{f_{\pi B}}{m_{\pi}}\right)^{2}\Pi_{\pi}$ &
$2\left(f_{\pi B}/m_{\pi}\right)^{2}\Pi_{\pi}$ &
\nodata \\
\enddata
\label{tab:BSE}
\tablecomments{
The index $i$ is specified in the left column, where $V (T)$ stands for the vector (tensor) coupling at each meson-$BB^\prime$ vertex.
The last row is for the (pseudovector) pion contribution.
}
\end{deluxetable}
%%%%%%%%%%%%%%%%%%%%%%%%%%%%%%%%%%%%%%%%%%%%%%%%%%%%%%%%%%%%%%%%%%%%%%%%%%%%%%%

%%%%%%%%%%%%%%%%%%%%%%%%%%%%%%%%%%%%%%%%%%%%%%%%%%%%%%%%%%%%%%%%%%%%%%%%%%%%%%%
\begin{deluxetable}{lcccc}
% \tabletypesize{\footnotesize}
\tablewidth{0pt}
\tablecaption{
Coupling constants related to the nucleon.
}
\tablehead{
\colhead{Symmetry} & \colhead{$g_{\sigma N}^{2}/4\pi$} & \colhead{$g_{\omega N}^{2}/4\pi$} & \colhead{$g_{\phi N}^{2}/4\pi$} & \colhead{$g_{\rho N}^{2}/4\pi$} }
\startdata
SU(6)              & 3.78                              & 5.74                              & \nodata                         & 0.350                           \\
SU(3)              & 3.38                              & 5.41                              & 0.631                           & 0.412                           \\
\enddata
\label{tab:CCN}
\tablecomments{
The $\phi$ meson does not couple to the nucleon ($N$) in SU(6) symmetry, while it couples to $N$ in SU(3) symmetry.
}
\end{deluxetable}
%%%%%%%%%%%%%%%%%%%%%%%%%%%%%%%%%%%%%%%%%%%%%%%%%%%%%%%%%%%%%%%%%%%%%%%%%%%%%%%

%%%%%%%%%%%%%%%%%%%%%%%%%%%%%%%%%%%%%%%%%%%%%%%%%%%%%%%%%%%%%%%%%%%%%%%%%%%%%%%
\begin{deluxetable}{lcccccccc}
\tabletypesize{\small}
\tablewidth{0pt}
\tablecaption{
Properties of symmetric nuclear matter at $n_{0}$.
}
\tablehead{
\colhead{}                     &
\colhead{$M_{N}^{\ast}/M_{N}$} &
\colhead{$K_{0}$}              &
\colhead{$J_{0}$}              &
\colhead{$E_{\rm sym}$}        &
\colhead{$L_{0}$}              &
\colhead{$K_{\rm sym}$}        &
\colhead{$K_{\rm asy}$}        &
\colhead{$K_{{\rm sat},2}$}    \\
\colhead{Symmetry}             &
\colhead{}                     &
\colhead{(MeV)}                &
\colhead{(MeV)}                &
\colhead{(MeV)}                &
\colhead{(MeV)}                &
\colhead{(MeV)}                &
\colhead{(MeV)}                &
\colhead{(MeV)}                }
\startdata
SU(6) & 0.742 & 275 & $-366$ & 32.5 & 75.3 & $-48.0$ & $-500$ & $-400$ \\
SU(3) & 0.747 & 269 & $-364$ & 32.5 & 78.0 & $-44.5$ & $-513$ & $-407$ \\
\enddata
\label{tab:nuclear-properties}
\tablecomments{
The effective mass of nucleon, incompressibility, third-order incompressibility, and symmetry energy are respectively denoted by $M_{N}^{\ast}$, $K_{0}$, $J_{0}$, and $E_{\rm sym}$.
The slope parameter and curvature parameter of the symmetry energy, $L_{0}$ and $K_{\rm sym}$, are also listed.
Using the parabolic approximation for the EoS, the 2nd derivative of the isobaric incompressibility coefficient is given by $K_{{\rm sat},2}=K_{\rm asy}-\frac{J_{0}}{K_{0}}L_{0}$ with the parameter $K_{\rm asy}=K_{\rm sym}-6L_{0}$ \citep{Chen:2009wv,Agrawal:2012rx,Sulaksono:2012ny}.
}
\end{deluxetable}
%%%%%%%%%%%%%%%%%%%%%%%%%%%%%%%%%%%%%%%%%%%%%%%%%%%%%%%%%%%%%%%%%%%%%%%%%%%%%%%

%%%%%%%%%%%%%%%%%%%%%%%%%%%%%%%%%%%%%%%%%%%%%%%%%%%%%%%%%%%%%%%%%%%%%%%%%%%%%%%
\begin{deluxetable}{lcccc}
% \tabletypesize{\footnotesize}
\tablewidth{0pt}
\tablecaption{
Coupling constants related to the hyperons.
}
\tablehead{
\colhead{Set}              & \hspace*{3.5cm} & \colhead{ESC08Y} & \hspace*{3.5cm} & \colhead{SU(3)Y}                  }
\startdata
$g_{\sigma\Lambda}$        & \               &  2.48            &                 &  1.74                             \\
$g_{\sigma\Sigma}$         & \               &  1.98            &                 &  1.17                             \\
$g_{\sigma\Xi}$            & \               &  1.95            &                 &  1.35                             \\
$g_{\sigma^{\ast}\Lambda}$ & \               &  0.876           &                 &  \nodata\!\!\!\!\tablenotemark{a} \\%0.000\tablenotemark{a} \\
$g_{\sigma^{\ast}\Sigma}$  & \               &  0.876           &                 &  \nodata\!\!\!\!\tablenotemark{a} \\%0.000\tablenotemark{a} \\
$g_{\sigma^{\ast}\Xi}$     & \               &  1.05            &                 &  \nodata\!\!\!\!\tablenotemark{a} \\%0.000\tablenotemark{a} \\
$g_{\omega\Lambda}$        & \               &  2.82            &                 &  1.85                             \\
$g_{\omega\Sigma}$         & \               &  2.82            &                 &  1.85                             \\
$g_{\omega\Xi}$            & \               &  2.09            &                 &  1.37                             \\
$g_{\phi\Lambda}$          & \               & -2.16            &                 & -1.42                             \\
$g_{\phi\Sigma}$           & \               & -2.16            &                 & -1.42                             \\
$g_{\phi\Xi}$              & \               & -3.11            &                 & -2.04                             \\
% $g_{\rho\Lambda}$          & \               &  0               &                 &  0                                \\
$g_{\rho\Sigma}$           & \               &  1.38            &                 &  1.28                             \\
$g_{\rho\Xi}$              & \               &  0.692           &                 &  0.642                            \\
\enddata
\label{tab:CCY}
\tablecomments{
All the values are divided by $\sqrt{4\pi}$.
We list the original set suggested by the ESC08 model (ESC08Y) and the calculated set based on naive SU(3) symmetry (SU(3)Y).
We assume $g_{\sigma^{\ast}\Lambda}=g_{\sigma^{\ast}\Sigma}$ and $g_{\rho\Lambda}=0$.
}
\tablenotetext{a}{
Because the $\sigma$-meson contribution in SU(3)Y already gives $U_{\Lambda}^{(\Xi)}=-19$ MeV and $U_{\Xi}^{(\Xi)}=-15$ MeV at $n_{0}$, the additional, attractive force due to the $\sigma^{\ast}$ meson is not required.
}
\end{deluxetable}
%%%%%%%%%%%%%%%%%%%%%%%%%%%%%%%%%%%%%%%%%%%%%%%%%%%%%%%%%%%%%%%%%%%%%%%%%%%%%%%

%%%%%%%%%%%%%%%%%%%%%%%%%%%%%%%%%%%%%%%%%%%%%%%%%%%%%%%%%%%%%%%%%%%%%%%%%%%%%%%
\begin{deluxetable}{lccccccccc}
\tablecolumns{9}
% \tabletypesize{\footnotesize}
\tablewidth{0pt}
\tablecaption{
Phase transition properties at the critical density, $n_{B}^{(c)}$ (fm$^{-3}$).
}
\tablehead{
\colhead{}     & \hspace*{2.0cm} & \multicolumn{3}{c}{ESC08Y} & \hspace*{2.0cm} & \multicolumn{3}{c}{SU(3)Y} \\
\cline{3-5} \cline{7-9} \\
\colhead{Case ($\beta$)}  &
\colhead{}                &
\colhead{$n_{B}^{(c)}$}   &
\colhead{$B^{(c)}$}       &
\colhead{$\mu_{B}^{(c)}$} &
\colhead{}                &
\colhead{$n_{B}^{(c)}$}   &
\colhead{$B^{(c)}$}       &
\colhead{$\mu_{B}^{(c)}$} }
\startdata
(1) 0.000      &                 & 1.065  & 400 & 1702        &                 & 1.775 & 400 & 2031         \\
(2) 0.025      &                 & 0.820  & 224 & 1474        &                 & 1.185 & 131 & 1605         \\
(3) 0.050      &                 & 0.680  & 184 & 1350        &                 & 0.945 & 105 & 1444         \\
(4) 0.100      &                 & 0.565  & 143 & 1246        &                 & 0.660 & 107 & 1274         \\
(5) 0.150      &                 & 0.500  & 123 & 1188        &                 & 0.505 & 121 & 1188         \\
(6) 0.200      &                 & 0.450  & 115 & 1147        &                 & 0.450 & 115 & 1147         \\
\enddata
\label{tab:Phase-transition-properties}
\tablecomments{
We list six cases, where the parameter $\beta$ varies between 0 to 0.2, for ESC08Y and SU(3)Y.
The critical bag constant, $B^{(c)}$, and the critical chemical potential, $\mu_{B}^{(c)}$, which are in the unit of MeV, are calculated by Equation \eqref{eq:DDbag}.
}
\end{deluxetable}
%%%%%%%%%%%%%%%%%%%%%%%%%%%%%%%%%%%%%%%%%%%%%%%%%%%%%%%%%%%%%%%%%%%%%%%%%%%%%%%

%%%%%%%%%%%%%%%%%%%%%%%%%%%%%%%%%%%%%%%%%%%%%%%%%%%%%%%%%%%%%%%%%%%%%%%%%%%%%%%
\begin{deluxetable}{lccccccccc}
\tablecolumns{9}
% \tabletypesize{\footnotesize}
\tablewidth{0pt}
\tablecaption{
Properties of a neutron star in ESC08Y and SU(3)Y.
}
\tablehead{
\colhead{}     & \hspace*{1.0cm} & \multicolumn{3}{c}{ESC08Y} & \hspace*{1.0cm} & \multicolumn{3}{c}{SU(3)Y} \\
\cline{3-5} \cline{7-9} \\
\colhead{Case ($\beta$)}       &
\colhead{}                     &
\colhead{$R_{\max}$}           &
\colhead{$M_{\max}/M_{\odot}$} &
\colhead{$n_{c}$}              &
\colhead{}                     &
\colhead{$R_{\max}$}           &
\colhead{$M_{\max}/M_{\odot}$} &
\colhead{$n_{c}$}              }
\startdata
(1) 0.000      &                 & 11.13 & 2.029 & 1.035      &                 & 11.36 & 1.874 & 1.020      \\
(2) 0.025      &                 & 11.56 & 2.003 & 1.005      &                 & 11.36 & 1.874 & 1.020      \\
(3) 0.050      &                 & 11.71 & 1.958 & 0.975      &                 & 11.46 & 1.873 & 1.025      \\
(4) 0.100      &                 & 11.63 & 1.896 & 1.010      &                 & 11.51 & 1.863 & 1.030      \\
(5) 0.150      &                 & 11.50 & 1.866 & 1.050      &                 & 11.43 & 1.856 & 1.060      \\
(6) 0.200      &                 & 11.41 & 1.853 & 1.070      &                 & 11.37 & 1.850 & 1.075      \\
\enddata
\label{tab:NS-properties}
\tablecomments{
We list the neutron-star radius, $R_{\max}$ (in km), the ratio of the neutron-star mass to the solar mass, $M_{\max}/M_{\odot}$, and the central density, $n_{c}$ (in fm$^{-3}$), at the maximum-mass point.
}
\end{deluxetable}
%%%%%%%%%%%%%%%%%%%%%%%%%%%%%%%%%%%%%%%%%%%%%%%%%%%%%%%%%%%%%%%%%%%%%%%%%%%%%%%

\end{document}